\documentclass[12pt,preprint]{aastex}
\def\pp{\parshape 2 0.0truecm 16.25truecm 2truecm 14.25truecm}
\newcommand{\be}{\begin{equation}}
\newcommand{\ee}{\end{equation}}

\def\lta{\,\raise 0.3 ex\hbox{$ < $}\kern -0.75 em
 \lower 0.7 ex\hbox{$\sim$}\,}
\def\gta{\,\raise 0.3 ex\hbox{$ > $}\kern -0.75 em
 \lower 0.7 ex\hbox{$\sim$}\,} 
\begin{document}

\title{Self-Similar Collapse Solutions for Cylindrical Cloud Geometries
\\ and Dynamic Equations of State} 

\author{Lisa Holden,$^1$ Kevin Hoppins,$^{1}$, 
Benjamin Baxter$^1$ and Marco Fatuzzo$^2$} 
 
\affil{$^1$Mathematics Department, Northern Kentucky University, Highland Heights, KY, 41099} 

\affil{$^2$Physics Department, Xavier University, Cincinnati, OH 45207}

\email{holdenl@nku.edu, fatuzzo@xavier.edu}

\begin{abstract} 

A self-similar formalism for the study of the gravitational collapse
of molecular gas provides an important theoretical framework
from which to explore the dynamics of star formation.  Motivated by 
the presence of elongated and filamentary structures observed
in giant molecular clouds, we build upon the existing body of
work on cylindrical self-similar collapse flows by including 
dynamic equations of state that are different from the
effective equation of state that produces the initial density distribution.
We focus primarily on the collapse of initial states for which the gas
is at rest and everywhere overdense from its corresponding
hydrostatic equilibrium profile by a factor $\Lambda$, and
apply our results toward the analysis of star formation within 
dense, elongated molecular cores.   
An important aspect of this work is the determination of the 
mass infall rates over a range of the parameters which define
the overall state of the gas -- the overdensity parameter
$\Lambda$, the index $\Gamma$ of the static equation of state, 
and the index $\gamma$ of the dynamic equation of state.
While most of the parameter space explored in this work
leads to solutions for which the underlying equations do not
become singular, we do include a discussion on how to 
treat cases for which solutions pass smoothly through the singular 
surface. In addition, we also present a different class of collapse solutions
for the special case $\gamma = 1$.   
\end{abstract}




\section{INTRODUCTION} 

Star formation in our galaxy occurs primarily within giant
molecular clouds (GMC) - highly nonuniform complexes of molecular gas
containing a total mass of $\sim 10^5\;M_\odot$ within
a radius of $\sim 20$ pc. These complexes have hierarchical structure
that can be characterized in terms of clumps and dense cores
surrounded by an interclump gas of density $\sim 5$--$25$ cm$^{-3}$.
Clumps have characteristic densities of
$\sim 10^3$ cm$^{-3}$ and radii ranging between $0.2$--$2$ pc,
the largest of which are comprised of as many as
$\sim 1000$ small ($R \sim 0.1 - 0.2$ pc), dense ($\sim 10^4$--$10^5$ cm$^{-3}$)
cores whose mass function has been measured to range from $\sim 1 - 100 M_\odot$
(with a peak $\sim 10 M_\odot$) by Jijina et al. (1999), and more recently,
from $\sim 0.2 - 20 M_\odot$ (with a characteristic mass of 
$\approx 2 M_\odot$) by Lada et al. (2008).  It is the gravitational collapse
of these cores within a clump that results in the formation of stars.
The dynamics of core collapse is therefore a fundamental component of
the star formation process.  

The study of self-similar collapse flows has provided 
an important cornerstone of the current theory of star formation (e.g., Shu et al.
1987), with numerous works appearing in the literature. 
The original self-similar collapse calculations (Larson 1969ab;
Penston 1969ab; Shu 1977) considered isothermal spherical flows.
Since then, many generalizations of the collapse have been made. The
leading order effects of rotation have been studied, both for the
inner pressure free region (Ulrich 1976; Cassen \& Moosman 1981) and
for the entire core (Terebey, Shu, \& Cassen 1984).  
The leading order effects of
magnetic fields have also been included (Galli \& Shu 1993ab; Li \&
Shu 1996, 1997). More recently, the collapse of magnetized singular
isothermal toroids has been studied (Allen, Shu, \& Li 2003; Allen,
Li, \& Shu 2003).

While much of the focus has been given to spherically symmetric flows, 
filamentary and elongated structures in molecular
clouds are commonly observed (e.g., Houlahan \& Scalo 1992; Harjunpaa et al. 1999;
Jijina 1999), and based on the results of numerical simulations 
(e.g., Curry 2000; Jappsen et al 2005), 
appear to be an important aspect of the star formation process.  
A complete understanding of molecular cloud dynamics
thus requires the inclusion of cylindrical geometries. 
Indeed, several authors have applied self-similar 
techniques toward the study of how cylindrical structures collapse
(Inutsuka \& Miyama 1992; Kawachi \& Hanawa 1998;
Hennebelle 2003; Tilley \& Pudritz 2003;  Shadmehri 2005).   
A central element of these studies involves the form of the equation of 
state used.  While an isothermal equation of state clearly provides a reasonable and important
starting point for the general study of core collapse, observational 
evidence points to softer equations of state.  Specifically, 
non-thermal linewidths $\Delta v$ in molecular cloud cores
show a correlation with density of the form $(\Delta v)^2 \sim
\rho^{-\beta}$, with $\beta > 0$ (e.g., Larson 1981; Jijina et
al. 1999).  If one interprets the linewidth $\Delta v$ as the
effective transport speed in the medium, then the corresponding
effective equation of state has the form $P \sim \rho^\Gamma$, where
$\Gamma = {1-\beta}$ and hence $0 < \Gamma \le 1$. 

The self-similar collapse of filamentary structures with polytropic equations of state
has been explored by Kawachi \& Hanawa (1998) and Shadmehri (2005),
with the latter work also including the effects of magnetic fields.
These works, however, considered static equations of state which imposed
global constraints on the pressure and density.
It is likely that the physical processes which govern a gas will change
as that gas undergoes gravitational collapse.  We
therefore extend the analyses of previous works
on cylindrical flows by considering a dynamic equation of state
(Fatuzzo, Adams, \& Myers 2004).
Specifically, we consider the case in which the dynamic
equation of state for the collapsing gas is different than the
effective (static) equation of state that produces the initial
equilibrium configuration. 
Here, the static equation of state (as set by $\Gamma$) refers
to the pressure law that enforces the initial (pre-collapse)
configuration for the gas, whereas the dynamic equation of
state (as set by $\gamma$) refers to the pressure law that describes
how the thermodynamic variables of the gas change as the material is
compressed during collapse. This process is governed by the entropy
evolution equation (introduced in \S 2.1), so that specific
entropy is conserved along a given streamline. 
However,  since the physics that determines the density profiles of the pre-collapse
states can be different from the physics that governs the
thermodynamics of the collapse flow, we allow $\gamma \ne
\Gamma$. Our analysis therefore allows for a more robust model 
from which to gain insight into 
the dynamics of filamentary/elongated molecular structures.  

An important aspect of this work is its focus on the mass
infall rate $\dot M$.  Collapse flows are
often self-similar and have no characteristic mass scale. 
However, the mass infall rate is one of the most important physical
quantities in the star formation problem, and determines, in
part, the total system luminosity and the total column density of the
infalling envelope.  These quantities, in turn, largely account for
the spectral appearance of protostellar objects (e.g., Adams, Lada, \&
Shu 1987; Adams 1990) since most of the
luminosity is derived from material falling through the gravitational
potential well of the star. Although the circumstellar disk stores
some of the energy in rotational motion, the system luminosity is
(usually) a substantial fraction of the total available luminosity 
\be 
L_0 \equiv {G M_s {\dot M} \over R_\ast } \, , 
\label{eq:luminosity} 
\ee 
where $M_s$ is the total mass of the star/disk system, ${\dot M}$ is the mass
infall rate, and $R_\ast$ is the stellar radius. The stellar radius,
which helps determine the depth of the potential well, is itself a
function of the mass infall rate (Stahler, Shu, \& Taam 1980).

The paper is organized as follows.  We formulate the collapse problem
via self-similar methods for the general case
of a collapsing cylindrical cloud of gas in \S2. In \S3, we determine 
the range of parameter space that yields collapse solutions
by considering the limiting cases $v/x \to 0$ and
$x/v \to 0$, and illustrate how the ensuing collapse is affected
by the initial state of the gas. 
Guided by these results, we explore in \S4 the collapse of 
initial states that are out of
exact hydrostatic equilibrium by being overdense, and
apply our results to the collapse of elongated cores in \S 5.  
We consider cases for which solutions go smoothly 
through the singular surface in \S6, including the $\gamma = 1$
case which leads to a different class of collapse solutions.
We present our conclusions in \S7.  

\section{FORMULATION OF THE COLLAPSE PROBLEM}  

\subsection{Basic Governing Equations}  

We adopt a cylindrical coordinate system described by the
variables $r, z$ and $\phi$.  
To keep the problem tractable, we assume dependence only in 
$r$ (i.e., the cylindrical structure is infinite in length),
with the self-gravitating gas
described locally by its density $\rho(r,t)$, pressure
$P(r,t)$ and (radial) velocity
$u(r,t)$, and globally by the mass per unit length $M_L(r,t)$ 
contained within a radius $r$.  The gravitational collapse of 
this fluid is governed by conservation of mass 
\be
{\partial M_L \over \partial t} + u {\partial M_L \over \partial r} = 0  
\qquad {\rm and} \qquad 
{\partial M_L \over \partial r} = 2 \pi r \rho , 
\ee
or equivalently, by the equation of 
continuity, 
\be
{\partial \rho \over \partial t} + {1 \over r}
{\partial \over \partial r} (r \rho u) = 0 \, ,  
\label{eq:fullcont} 
\ee
and the force equation 
\be
{\partial u \over \partial t} + u
{\partial u \over \partial r} = - {1 \over \rho} 
{\partial P \over \partial r} - {2 G M_L \over r} \, .  
\ee 

To complete the description required for the evolution 
of the gas to be solved, the pressure must be specified
through a choice of the equation of state.  For example, 
the relation $P = s^2 \rho$ is used to describe an isothermal gas (where
$s$ is the sound speed), whereas a polytropic 
equation of state $P = {\cal K} \rho^\gamma$ allows for a more
general treatment of the problem.  Of course, adopting
an equation of state represents a simplification 
to the real system as it embodies the numerous
physical processes which govern the true state of the
gas -- and how they add/remove energy to/from the gas --  
into one simple relation.  It is quite likely that
these processes will change significantly during
the collapse, so that the equation of state which
governs the gas during the initial stages of collapse
will almost certainly have a different form from that
which governs the collapse during the later stages of the collapse.
We therefore introduce a dynamic polytropic
equation of state in our formalism which
allows the relation between pressure and density to evolve
during the collapse (e.g., Fatuzzo, Adams \& Myers 2004).
Specifically, we assume that entropy is conserved along
a streamline, and use the conservation of entropy 
equation  
\be
\left({\partial \over \partial t} + u {\partial \over \partial r}\right) 
 \log \left[P / \rho^\gamma \right]  \, = 0 \, , 
\label{eq:entropyone} 
\ee
to follow the evolution of the pressure.  We thus refer to 
$\gamma$ as the index of the dynamic equation of state.
It is important to note that equation (\ref{eq:entropyone}) 
describes how a given parcel of gas changes its
thermodynamic variables along a streamline, therefore allowing for
an equation of state which evolves during the collapse.  
In contrast, relating the pressure to the
density through a fixed (static)
equation of state (e.g., $P = {\cal K} \rho^\gamma$) implies a global constraint on those
variables.  

\subsection{The Similarity Transformation}  

As shown in \S2.1, the cylindrical collapse problem is
represented mathematically by a set of coupled partial 
differential equations in time $t$
and radial position $r$.
In this section, we find a similarity transformation that
reduces this set of PDE's to a set of ordinary differential equations in a
new similarity variable $x$ which we define below. In particular, we
look for a similarity transformation of the general form 
$$
x = A t^a r \, , \qquad 
\rho = B t^b \alpha(x) \, , \qquad 
M_L = C t^c m(x) \, , 
$$
\be
u = D t^d v(x) \, , \qquad {\rm and} \qquad 
P = E t^e p(x) \, . 
\ee
Here, both the coefficients ($A$, $B$, $C$, $D$, $E$) and the indices
$(a, b, c, d, e)$ are constants.  The reduced fluid fields ($\alpha$,
$m$, $v$, $p$) are dimensionless functions of the (single)
dimensionless similarity variable $x$. In this work, the time benchmark $t=0$ 
corresponds to the instant of the onset of collapse.

The general similarity transformation calculation leads 
to four equations to specify the five indices $a,b,c,d,e$.  
We leave the constant $a$ arbitrary for the moment and write 
the rest of the variables in terms of its value, i.e., 
$$
a=a \, , \qquad b = - 2 \, , \qquad c = - (2a + 2) \, , 
$$
\be
d= - (a + 1) \, , \qquad {\rm and} \qquad 
e = - 2 (a + 2) \, . \ee
Similarly, for the coefficients we obtain
$$
A = A \, , \qquad 
B = (2 \pi G)^{-1} \, , \qquad 
C = (A^2 G)^{-1} \, , 
$$
\be
D = A^{-1} \, , \qquad {\rm and} \qquad 
E = (2 \pi G A^2)^{-1} \, ,
\ee
where $G$ is the gravitational constant. 
We thus obtain reduced equations of motion in the form 
\be(ax + v) {dm \over dx} = (2a + 2)  m \, ,
\label{eq:contmass}
\ee
\be
{dm \over dx} = x \alpha \, ,
\label{eq:contmass2} 
\ee 
\be
(ax + v) {1 \over \alpha} {d \alpha \over dx} + 
{dv \over dx} = \left(2 - {v\over x}\right) \, ,
\label{eq:contrho} 
\ee

\be
(ax + v) {dv \over dx} + {1 \over \alpha} {dp \over dx} 
= - {2 m \over x} + (a+1) v \, , 
\label{eq:force} 
\ee 
\be
(ax + v) {d \over dx} \log [ p/\alpha^\gamma ] = 
2 (2 + a - \gamma) \, .
\label{eq:entropy} 
\ee 
Note that our similarity transformation is not unique -- one can always rescale
the coefficients $\{ A, B, C, D, E \}$ by a set of dimensionless
numbers and obtain new equations of motion with different numerical
coefficients.

The first two equations of
motion can be immediately combined to obtain 
an expression for the reduced mass $m(x)$, i.e., 
\be
m = {(ax + v) \over (2a + 2) } x \alpha \, . 
\label{eq:massint} 
\ee
Likewise, multiplying equation (\ref{eq:contmass}) by the constant 
\be
q \equiv {2 \over 2a + 2} (2 + a - \gamma) . 
\label{eq:qdef} 
\ee
and subtracting the product
from equation (\ref{eq:entropy}) yields a differential equation 
which can be integrated to obtain an expression for the reduced pressure
\be
p = \, {\cal C}_1 \, \alpha^{\gamma} \, m^q \, 
= {\cal C}_1 \, \alpha^{\gamma+q}\,
\left[{(ax+v)\over (2a+2)}\,x\right]^q\, , 
\label{eq:psolution} 
\ee
where ${\cal C}_1$ is a positive integration constant.

Given the solutions for
the reduced pressure $p(x)$ and reduced mass $m(x)$, Eqs. 
(\ref{eq:contrho}) and (\ref{eq:force}) are the relevant equations of
motion to determine the remaining unknown functions $\alpha(x)$ and
$v(x)$. Using Cramer's rule, we derive an equivalent set of
equations 
\be
{d\alpha\over dx} = {{\cal A}\over {\cal D}} 
\qquad {\rm and} \qquad
{dv \over dx} = {{\cal V}\over {\cal D}}\,,
\ee
where
\be
{\cal D} = {\left(ax+v\right)^2\over\alpha}-{\gamma p \over\alpha^2}\,,
\ee
\be
{\cal V} = -{\left(ax+v\right)^2\over 1+a}-{2p\left(2+a-\gamma\right)\over\alpha^2}
+\left(ax+v\right)\left(1+a\right){v\over\alpha}
-{\gamma p\over\alpha^2}\left(2-{v\over x}\right)\,,
\ee
and
\be
{\cal A} = \left(ax+v\right)\left(2-{v\over x}\right)+{\alpha\left(ax+v\right)\over
1+a}+{2p\left(2+a-\gamma\right)\over\alpha \left(ax+v\right)}
-(1+a)v\;.
\ee
The physical state of the gas is defined through a choice
of the parameters ($a, \gamma, C_1$), and its collapse is
described by the solutions to  Eqs. (14) -- (20)
for the specified set of reduced field variables  
$v_i = v(x_i)$ and $\alpha_i = \alpha(x_i)$.

It is
mathematically possible to consider ``complete'' solutions that
span the entire available range $-\infty < x < \infty$
(first obtained by Hunter 1977; see also
Whitworth \& Summers 1985), and as such, span
both negative and positive times. 
However, it is rather unlikely that molecular clouds will
evolve toward their centrally condensed initial configurations in a
self-similar manner subject to (only) the physics included in these
equations of motion.   For example, before the onset of collapse (for $t < 0$),
molecular clouds may evolve through the processes of ambipolar
diffusion (at least for small mass scales), shocks, turbulent
dissipation, cooling flows, condensation instabilities, and
cloud-cloud collisions.
In addition, the cloud will most likely
initiate collapse before a completely self-similar equilibrium state
has been attained; the collapse will only become self-similar
asymptotically in time (i.e., the self-similar collapse solutions of
this paper are intermediate asymptotic solutions to the realistic
problem of the collapse of a finite cloud with finite central
density). As such, the self-similar solutions of
this paper for the protostellar collapse phase ($t > 0$) cannot (in
general) be extended to the pre-stellar phase ($t < 0$), as they would
likely encounter a critical point and become singular. 
We therefore limit this discussion to solutions with $0
< x < \infty$, sometimes called ``semi-complete'' solutions.

\section{GENERAL SOLUTIONS TO THE COLLAPSE PROBLEM}

In order to determine what set of parameters ($a,\gamma, C_1$) 
yield solutions that describe the collapse of a molecular cloud
filament, we obtain analytic solutions to the equations of motion
in the limit that $t \to 0$.  As these solutions describe
the early stages of the collapse, the gas velocity $u$ must be
negative and bounded.  For a positive value of $a$, this constraint
requires that $u = D t^{-(a+1)} v(x)$ not be singular at $t = 0$,
which in turn requires that $v(x) \propto x^\beta \propto t^{\beta a}$, where
$\beta \ge (a+1)/a$. Clearly then, the ratio $v/x \to 0$ as $t \to 0$
for solutions relevant to our discussion.

In this limit,  Eq. (9) reduces to a form that can be easily integrated, and 
along with Eqs. (14) and (16), leads to analytic solutions for the
reduced density, mass, and pressure of the form
\be
\alpha = \lambda\, x^{2/a}\;,
\ee
\be
m = {a\over 2a+2}\, \lambda\, x^{(2a+2)/a}\;,
\ee
and
\be
p = C_1\,\lambda^{-aq}\,\left[{a\over 2a+2}\right]^q\,\alpha^{2+a} \,,
\ee
where $\lambda$ is a positive constant.  
The reduced velocity is
governed by the limiting form of Eq. (12)
\be
a {dv\over dx} - (a+1) {v\over x} = 
V_0\, x^{2/a}\,,
\ee
where 
\be
V_0 = -2\lambda\left[{a\over 2a+2}\right]\,\left[1 + C_1 \left({2+a \over a}\right) 
\left({2a+2\over a}\,\lambda^a\right)^{(\gamma-1)/(1+a)}\right]\,.
\ee
Eq. (24) yields a power-law form solution
\be
v = V_0\, x^{(2+a)/a}\,,
\ee
indicating that collapse solutions can be found for $a > 0$ (since $V_0 < 0$).
However, the real density becomes time-independant and scales with
radius as $\rho\propto r^{2/a}$ in the limit that
$x/v \to 0$.
Since the density approaches zero at small radii, solutions for $a > 0$
cannot represent the collapse of filamentary like structures (whose initial
density profiles are peaked at $r = 0$), thereby ruling 
out this range of parameter space in our work.

Interestingly, the ratio $v/x$ also approaches zero as $t \to 0$ 
for the case that $a < 0$ (since $x \to \infty$).
The reduced density, mass and pressure given by
Eqs. (21) -- (23) therefore also describe the state of the
gas at the onset of collapse for this case. However, the real
density (which still scales as $\rho\propto r^{2/a}$) now becomes singular as
$r \to 0$.  Of course, the equations of motion
presented in this work are simply mathematical idealizations
to the real, physical problem, for which filaments have a finite density 
core at small radii $r < r_C$ and a large but finite outer boundary $r_{out}$.  
This "real" system is expected to follow
the self-similar solutions for intermediate length-scales. Indeed, 
previous numerical work
(Foster \& Chevalier 1993) indicates that the collapse of an
isothermal core approaches the expected self-similar form when the
core has $r_{out}/r_C > 20$.  Likewise, Fatuzzo, Adams \& Myers
(2004) found similar results, with cores that have initial inward velocities
more readily approaching the self-similar collapse forms.

For the case that $a$ is negative, additional constraints are provided 
by the requirement that the reduced mass $m$
remain positive, and that the reduced pressure $p$ increases as $\alpha$ increases.
Together, these physical conditions require $-2 < a < -1$, as can be clearly
seen by the forms of Eqs. (22) and (23).  
In addition, collapse solutions (for which $V_0 < 0$) require that
\be
\lambda > \lambda_{crit} \equiv  \left[{a \over 2a+2}\right]^{1/a}\,
\left[-C_1{(2+a) \over a}\right]^{(a+1)/(a-a\gamma)}\,.
\ee
We note that $\lambda_{crit}$ is not defined if $\gamma = 1$. 
Indeed, the $\gamma = 1$ case yields a unique class of collapse
solutions, and will be considered separately in \S 6.2.  

Since the real pressure
is defined in terms of the constant $A$ through the
similarity transforms, we can set $p = \alpha^{a+2}$ 
without loss of generality.  That is, for a real pressure
\be
P = {\cal K} \rho^\Gamma\,,
\ee
setting  
\be
A = \left[{\cal K} (2\pi G)^{1-\Gamma}\right]^{-1/2}\,, 
\ee
and $\Gamma = (2+a)$
yields the desired form for the reduced equation of state
($p = \alpha^{a+2}$), and sets the integration constant
\be
C_1 = \left[{2a+2\over a}\right]^q \lambda^{aq}\,,
\ee
and, in turn, 
\be
\lambda_{crit} = \left[-{a^2\over (2a+2)(2+a)}\right]^{1/a}\,,
\ee

To obtain full solutions, one must numerically integrate Eqs. (17) -- (20)
from a specified initial set of values $\alpha_i = 
\alpha(x_i)$ and $v_i = v(x_i)$.  We note that one cannot set initial 
conditions which correspond to the onset of collapse since 
$x\to \infty$ as $t\to 0$. As such, initial conditions are set
through an arbitrary choice of $x_i >> 1$ along with 
a corresponding value of $\alpha_i$ for which
$\lambda > \lambda_{crit}$ and a small but negative value of $v_i$. 
Representative solutions are illustrated in Figs. (1) and (2)
for the case that $a = -1.25$ ($\Gamma = 0.75$) and $\gamma = 0.25$ (with
$C_1$ given by Eq. [30]), obtained by integrating inward from 
$x_i = 5 \times 10^3$ and $\alpha_i
= 5.78 \times 10^{-7}$ (for which $\lambda = 1.5 \lambda_{crit}$).  
The solid curves represent
the solutions obtained by setting the value of $v_i$ equal to
its power-law counterpart, as defined by Eq. (26).  The dashed
lines represent solutions obtained by setting the value of $v_i$ equal to
0.1, 0.3, 3 and 10 times the power-law value.  It is clear from our results that
the reduced density is not sensitive to the initial velocity, as
all five cases yield values that differed by less than $0.1$ percent. 
In addition, it is clear that the velocity evolves toward its power-law
form as given by Eq. (26). 

For completeness, we note that in the limit
that $x/v \to 0$, the reduced mass
approaches a constant value -- $m \to m_0$.  In 
turn, the reduced pressure takes the form $p \propto \alpha^\gamma$.  
Taken together, the two limiting solutions presented here thus illustrate how the
equation of state smoothly transforms from an initial index of $\Gamma = a+2$ (at
early times and/or large radii) to a dynamic index $\gamma$ (at late times
and/or small radii).  This result is clearly illustrated in Fig. 3, which plots $p$
versus $\alpha$ for the solutions presented in Figs. (1) and (2).  The 
five initial velocities yield results that are within $0.5$ percent of each other. 
As can be easily seen from Fig. (3), the equation of state index evolves from a 
value of $\Gamma = 0.75$
(in the lower left corner of the figure) to a value $\gamma = 0.25$ (in the
upper right corner), with the 
transition occurring within a narrow region around $x \approx 1$.

Finally, we note that the equations of motion which define our problem 
may well yield several different classes of solutions -- both physical (e.g.,
wind solutions) and unphysical.  However, solutions which 
pertain to the collapse of filamentary structures in the absence of
shocks (but see \S 6) are all qualitatively similar, and therefore characterized by
solutions such as those presented in Figs. (1) -- (3).   

\section{COLLAPSE OF OVERDENSE STATES} 

The analysis presented in the previous section illustrates
that collapse solutions of filamentary structures 
exist when
$-2 < a < -1$ and $\lambda > \lambda_{crit}$.  It is easy to show that
when $\lambda = \lambda_{crit}$, Eqs. (21) -- (23) 
yield self similar solutions which describe a gas in hydrostatic equilibrium
(for which $v = 0$).  
These solutions
are fully specified by the equation of state index $\Gamma$ 
(which we use hereafter to specify the state of the gas 
instead of $a$), and reduce to
\be
\alpha_E = \left[\Gamma\,{(2-2\Gamma)\over (2-\Gamma)^2}\right]^{1\over 2-\Gamma}
\,x^{-2\over 2-\Gamma}\;,
\ee
\be
m_E = \left[\Gamma\,{(2-2\Gamma)^{\Gamma-1}\over (2-\Gamma)^\Gamma}
\right]^{1\over 2-\Gamma}\, x^{2-2\Gamma \over 2-\Gamma}\,,
\ee
and $p = \alpha^\Gamma$ (assuming that $A$ is set through the relation given 
by Eq. [29]).

In turn, the real density and mass per unit length profiles 
are given by the time independent expressions
\be
\rho_E = \left[{{\cal K}\over 2\pi G}\right]^{{1\over 2-\Gamma}} \,
\left[\Gamma \,{(2-2\Gamma) \over (2-\Gamma)^2} \right]^{1\over 2-\Gamma}
\, r^{{-2\over 2-\Gamma}}\;,
\ee
and
\be
M_{L,E} = \left[{\cal K} (2\pi G)^{1-\Gamma}\right]^{{1\over 2-\Gamma}} \,
\,G^{-1}\,
\left[\Gamma \,{(2-2\Gamma)^{\Gamma-1}\over (2-\Gamma)^\Gamma}
\right]^{1\over 2-\Gamma}\, r^{2-2\Gamma \over 2-\Gamma}\,.
\ee

In this section, we 
consider the collapse of gas structures which are initially
at rest, but are out of hydrostatic equilibrium by being 
everywhere overdense by a factor $\Lambda$.  As such, the initial
gas (at time $t = 0$) is described by 
\be
\alpha_i = \Lambda \alpha_E
\qquad {\rm and} \qquad
v_i = 0\;,
\ee
where $\alpha_E$ is the corresponding density profile if the gas were in
hydrostatic equilibrium.  Essentially, this class of solutions
correspond to the solutions discussed in \S 3, where 
$\lambda = \Lambda \lambda_{crit}$, but with the 
velocity in the $x \to \infty$ limit given by Eq. (26), 
as shown below.

We note that hydrostatic equilibrium solutions of the form
\be
\alpha_E = \bar\Lambda x^{2/a}
\qquad {\rm and} \qquad
m_E = \bar\Lambda\, {a\over a+2}\, x^{(2+2a)/a}\,,
\ee
exist when $\gamma = 1$.  However, since the reduced
pressure takes the form $p = C_1
\alpha m$, the corresponding balance between 
the pressure gradient and gravity, as described
by Eq. (12),
\be
C_1 {d\over dx} (\alpha m) = - 2 {\alpha m \over x}\,,
\ee
is then always maintained regardless of the value of $\bar
\Lambda$.  The collapse scenarios considered in this section 
therefore cannot occur if the dynamic index $\gamma = 1$, regardless of the
value of the static index $\Gamma$. 
For completeness, we explore a different class of collapse solutions
with $\gamma = 1$ in \S 6.2.

With an initial (overdense) state defined in terms of
$\Gamma$ and $\Lambda$, as per Eqs. (32) and (36),
and the dynamic equation of state defined in terms of
$\gamma$, the ensuing collapse solutions can then be obtained 
by numerically integrating Eqs. (17).  
The adopted relation $p = \alpha^\Gamma$ is maintained for the
initial (overdense) state by setting 
\be
C_1 = \Lambda^{{(\Gamma-2)(\gamma-\Gamma)\over 1-\Gamma}}
\left({2-\Gamma\over \Gamma}\right)^{{\gamma-\Gamma\over
1-\Gamma}}\,.
\ee

In practice, however, this initial state corresponds to
a time $t_i = 0$, and in turn,  an initial value of $x_i = \infty$.
We therefore first obtain higher-order analytical expressions for
the reduced density and velocity from those presented in \S 3 via a series expansion
of the reduced equations in the limit $x_0 >> 1$ .  Doing so yields the following values for
the reduced density and velocity:
\be
\alpha_0 = \Lambda\,\alpha_E (x_0)\,
\left[1-{\Gamma\over 2-\Gamma} \,\Delta_0 \, x_0^{-2\over 2-\Gamma}\;\right]\,,
\ee
and
\be
v_0 = -\Delta_0 \,x_0^{-\Gamma\over 2-\Gamma}\;
\left[1+{1\over 3}\left\{\Lambda \beta_0 +
{2\Gamma^3\over (2-\Gamma)^2} \,{1\over (\Lambda\beta_0)^{1-\Gamma}}
-{\Gamma\over 2-\Gamma}\,\Delta_0\right\}\,x_0^{-2\over 2-\Gamma}\right]\,,
\ee
where
\be
\beta_0 = \left[\Gamma {(2-2\Gamma) \over (2-\Gamma)^2}\right]^{1\over 2-\Gamma}\,,
\ee
and
\be
\Delta_0 = \Lambda\,\left({2-\Gamma\over 1-\Gamma}\right)\,\beta_0
\,\left[1-\Lambda^{-(2-\Gamma)}\right]\,.
\ee
Full solutions can then be easily obtained by 
numerically integrating inward from $x_0$.

We present solutions for the reduced density and velocity
in Figs. 4 -- 7 for an overdensity parameter of
$\Lambda = 1.5$ and several different sets of ($\gamma, \Gamma$). 
The reduced density clearly exhibits the broken power law profile 
typical of self-similar collapse solutions.  We note that the spectral
indices which characterize the reduced density solutions are not sensitive
to the choice of $\gamma$, and hence have similar forms when $x << 1$.
In contrast, these solutions depend sensitively on the value of $\Gamma$, both
in terms of shape when $x >> 1$ and in overall normalization. 
Similar behavior is also clearly observed for the reduced velocity.  

As noted in the introduction, self-similar collapse flows  
have no characteristic mass scale. Instead, the
collapse flow feeds material onto the central star/disk system at a
well-defined mass infall rate $\dot M$. In these flows, the infalling
material always approaches free-fall conditions on the inside (in the
limit $r \to 0$) and the reduced mass determines the size of the
infall rate.  Specifically,
$m(x) \rightarrow m_0$ as $x\rightarrow 0$, and in turn,
the mass accretion rate per unit length becomes
\be
\dot M_L = 2 (1-\Gamma) {\cal K}\, (2\pi G)^{1-\Gamma} \,G^{-1}\,m_0
\,t^{1-2\Gamma} \,.
\ee
Note that if $\Gamma = 0.5$, $\dot M_L$ is constant
in time, whereas softer (stiffer) equations of
state result in temporally increasing (decreasing)
mass accretion rates. 
In contrast, the mass accretion rate for spherically symmetric flows 
is given by the expression
\be
\dot M =  (4-3\Gamma)\, {\cal K}^{3/2} \,
(4\pi G)^{3(1-\Gamma)/2}\, G^{-1}\, m_0 \, t^{3(1-\Gamma)}\;,
\ee
and is constant in time for an initially isothermal gas
(Fatuzzo, Adams \& Myers 2004).

We plot the value of $m_0$ as a function of $\Gamma$ for $\Lambda = 1.5$ and
three different values of $\gamma$ in Fig. 8, and
as a function of $\Lambda$ for several sets
of ($\Gamma$, $\gamma$) in Fig. 9 (the data point in this figure at
$\Lambda = 1.08$ is discussed in \S 6.1).
As expected, larger infall rates
occur for initial states that are more overdense.
In addition, stiffer static equations of state result in larger mass
accretion rates.
In contrast, the dynamic equation of state (as defined
by $\gamma$) has little effect on the value of $m_0$.

The insensitivity of the collapse dynamics to 
the value of the dynamic index $\gamma$ should not be surprising given the
inside-out nature of the collapse (as shown explicitly in 
\S5).  Specifically, the break at $x \approx 1$ exhibited in the density
and velocity profiles shown in Figs. 4 -- 7 denotes 
the boundary between gas that is in some part supported by 
pressure ($x >> 1$) and gas that has lost that support 
and is therefore approaching free-fall ($x << 1$).  
In terms of real quantities, this boundary occurs at a radius
\be
r_B \approx { 1\over A t^a} = {\cal K}^{1/2}\, (2\pi G)^{(1-\Gamma)/2} \, t^{2-\Gamma}\,,
\ee
that moves outward in time.  Since the speed at which 
this boundary moves is governed by the static equation of state, 
the collapse dynamics therefore depend sensitively on $\Gamma$.
In contrast, the dynamic gas pressure has little effect
on the collapse dynamics when $x < 1$ ($r < r_B$), and hence, 
on the overall collapse.

\section{APPLICATION TO ELONGATED CORES}

Dense star-forming cores in molecular clouds are on average
slightly elongated, with some cores exhibiting aspect ratios
as large as 5 (Jijina et al. 1999).   
We apply our results to the evolution of fairly elongated cores,
making the highly idealized assumption that their collapse can 
be reasonably approximated by our formalism.  In reality, how such
cores collapse  likely falls within the limiting cases of
cylindrical symmetry (explored here) and spherical symmetry (explored
in Fatuzzo, Adams \& Myers 2004).

Interpreting the observed non-thermal linewidths $\Delta v$ in molecular
cores as the effective
transport speed in the medium, one finds 
\be
\Delta v = \left[{\partial P\over \partial \rho}\right]^{1/2}
 = \left[{\cal K}\, \Gamma \right]^{1/2} \,\rho^{-(1-\Gamma)/2}\,,
\ee
for an assumed polytropic equation of state $P = {\cal K} \rho^\Gamma$.
Adopting the fiducial values $\Gamma = 0.5$ and $\Delta v = 1$ km/s for a density of 5,000 cm$^{-3}$
(e.g., Larson 1981) then yields a value of ${\cal K} = 2.6$  g$^{1/2}$ cm$^{1/2}$ s$^{-2}$,
and in turn, density and velocity profiles given by
\be
\rho(r,t) = 2.4 \times 10^{-21} \,{\rm g}\, {\rm cm}^{-3} 
\left({t \over 1\, {\rm Myr}}\right)^{-2} \alpha(x)\,,
\ee
\be
u(r,t) = 2.3 \,{\rm km/s} \, 
\left({t \over 1 \,{\rm Myr}}\right)^{1/2} \, v(x)\,,
\ee
where 
\be
x = 0.42 \,\left({r\over 1 {\rm pc}}\right)\,
\left({t\over 1 \,{\rm Myr}}\right)^{-3/2}\,.
\ee
In Figs. 10 and 11, we plot the density and velocity 
profiles for the collapse
from an initial state defined by $\Lambda = 1.5$ and
$\Gamma = 0.5$, and a dynamics index $\gamma = 0.5$.
The solid curve in Fig.
10 depicts the initial density profile, and the
dotted curves represent the ensuing profiles 
(from top to bottom) at times $t = 10^4$, $10^5$, and $10^6$ yrs.
Likewise, the dotted curves in Fig. 11 show the velocity 
profiles (from bottom to top) at times $t = 10^4$, $10^5$, and $10^6$ yrs.
These results clearly illustrate the inside-out nature
of the collapse, with gas inside the transition boundary
\be
r_B \approx 2.4\, {\rm pc}\,
\left({t\over 1 \,{\rm Myr}}\right)^{3/2}\,,
\ee
falling inwardly away from the overlying gas layers.  

From the onset of collapse, it takes a time
$t_* \sim 10^5$ yrs for the formation of a 
young stellar object to occur, and an additional
$t_d \sim 5\times 10^5$ yrs
before stellar outflows disperse the surrounding
core material.  Thus, our model predicts that cores 
associated with young stellar objects would be characterized
by densities $\sim 5\times 10^4$ cm$^{-3}$ and
radii $\sim 0.1$ pc, in good agreement with observations
(Jijina et al. 1999).  
In addition, the mass accretion rate per unit length is 
\be
\dot M = {\cal K} \left({2\pi\over G}\right)^{1/2} m_0 = 
1.2\times 10^3 M_\odot \,{\rm Myr}^{-1}\, {\rm pc}^{-1}\, m_0\,.
\ee
An elongated
core with a length of $l = 0.5$ pc would accrete a mass of
$M_{acc} \approx 10 M_\odot$ in a time $t_*$ for a slightly 
overdense initial state (since $m_0 \approx 0.15$ in this case,
as can be from the solid curve in Fig. 9),
and $M_{acc} \approx 37 M_\odot$ for $\Lambda = 1.5$
($m_0 = 0.6$).  These
results are consistent with the low star
formation efficiencies of $2 - 10$ \% deduced from 
observations, but seem to favor initial states that are only
slightly overdense. 

\section{SMOOTH SINGULAR SOLUTIONS}
\subsection{General Formalism}
The solutions presented in \S 3 did not cross
through the singular surface, defined in the three-variable space
$x, \alpha,$ and $v$ as the surface on which ${\cal D} = 0$. 
While several different forms of solutions
can pass through the singular surface (see, e.g., Lou \& Cao 2007),
we focus here on those solutions which pass through it smoothly.
That is, we consider solutions for which the derivatives
$v' = dv/dx$ and $\alpha' = d\alpha/dx$ exist on the singular surface. 
A necessary but not sufficient condition for the existence
of these solutions is that ${\cal V}$ (and hence ${\cal A}$) also 
become zero, which occurs along a unique curve (referred to 
as the critical curve) on the singular surface for each set
of values ($a, C_1, \gamma$).  Alternatively, for solutions which 
represent the collapse from an initially static, overdense 
state for which $p = \alpha^\Gamma$, a unique critical 
curve exists for each set of 
($\Lambda, \Gamma, \gamma$) values.  As a matter of illustration, 
we plot the ($v_c, x_c$) and ($\alpha_c, x_c$)
projections of the critical curve in Figure 12 for the parameters
$\Lambda = 1.5$, $\Gamma = 0.5$, $\gamma = 0.5$,
with
the upper panel presenting the reduced density profile, and the
lower panel presenting the reduced velocity profile (dotted curves).
The solid curves represent the solution for the 
corresponding collapse dynamics, which clearly does not 
cross the critical curve (or the singular surface).

In order to properly treat solutions which pass through the singular surface
smoothly, we obtain analytical solutions through a Taylor series
expansion about the
critical curve of the form  $x = x_c + \delta$, 
$v = v_c + \delta v_1$, and $\alpha = \alpha_c + \delta \alpha_1$,
where $\delta << 1$, and $v_1$ and $\alpha_1$ are the first
derivatives in $v$ and $\alpha$ evaluated on the critical curve.
Since the system ODE's are second order, only the first two terms
in the expansion are required.
Values for $v_1$ and $\alpha_1$ are obtained by substituting these
terms into equations (17) and expanding in
$\delta$.  Doing so then yields the
relations
\be
\alpha_1 = \left(2-{v_c\over x_c} - v_1\right)\,{\alpha_c\over
v_c-(2-\Gamma)x_c} \,,
\ee
and 
\be
-(1+\gamma) v_1^2 + \left[4\Gamma-1+2(1-\gamma){v_c\over x_c}\right] v_1 + C_v = 0\,,
\ee
where
\be
C_v = {\alpha_c\over 1-\Gamma}\left(\Gamma-{v_c\over x_c}\right)
+2 \left(\Gamma+1-{\Gamma\over\gamma}\right){v_c\over x_c}
-\gamma {v_c^2\over x_c^2} - {2\Gamma^2\over \gamma}\,.
\ee
The first derivative $v_1$ is then the real, positive solution to the above 
quadratic equation. Full solutions can then be obtained by numerically
integrating inward ($\delta < 0$) and outward ($\delta > 0$) 
from the corresponding Taylor series solutions. 

As an example, we consider the collapse from an initial state defined
by the parameters $\Lambda = 1.08$, $\Gamma = 0.25$ and $\gamma = 0.5$. Numerically
integrating Eqs. (17) inward from an initial value of $x_0 = 10^4$ (as
described in \S4), one finds that the equations become singular as
$x\rightarrow 0.336$, i.e., as the solutions approach the singular
surface.  Since a real and positive solution $v_1$ exists for 
Eq. (54) at $x_c = 0.336$ for this case, this solution can thus
pass smoothly through the singular surface (by crossing through the 
critical curve), and can be matched to the Taylor series
solution on the other side of the critical curve.  Further numerical 
integration shows that this solution also crosses smoothly through the
singular surface at a second point $x_c = 0.0322$, as illustrated 
in Figure 13.  The solid curve in the upper (lower) panel 
presents the reduced density (velocity) profile obtained for our solution, 
and the dotted curves represent the ($\alpha_c, x_c$) and ($v_c, x_c$) projections
of the critical curve for the case being considered.  
It is interesting to note that $\alpha$
approaches $\alpha_c$ as $x \rightarrow 0$.  The value of $m_0$ was
also determined and is depicted in Fig. 9 by the solid square
data point, and is consistent with the extrapolation of the 
dot-dash curve corresponding to the same ($\Gamma$, $\gamma$) values.

\subsection{The $\gamma$ = 1 case}
As noted in \S4, density profiles that are out of hydrostatic equilibrium by being
everywhere overdense (i.e., are of the form $\alpha_i = \Lambda \alpha_E$, $v_i = 0$)
do not exist for the $\gamma = 1$ case, regardless of the form for the initial equation
of state as defined by $a$ and $C_1$.  This result is somewhat surprising
given the apparent insensitivity that the solutions presented in
\S4 have on the choice of $\gamma$, but is borne out through numerical
investigation.  However, a different class of collapse 
solutions which passes smoothly
through the singular surface does exist for this case, and is presented 
here for completeness.  A full analysis of all solutions with 
$\gamma = 1$ (e.g., wind solutions, shock solutions) is outside the 
scope of this work and 
will be presented elsewhere.   

To keep our analysis as general as possible, we describe the
gas in terms of $a$ and $C_1$ throughout this
subsection.  It is easy to show that when $\gamma = 1$, 
the critical curve on the singular surface 
becomes a straight line defined by the linear relation $v_c = k x_c$ and 
a constant reduced density
\be
\alpha_c = {(2+2a)(a+k)\over C_1}\,,
\ee
where $k$ is a (real) solution to the quadratic equation
\be
\left[{2\over C_1}-1\right] k^2 + \left[3+{4a\over C_1}\right] k
+\left[4a+2a^2+{2a^2\over C_1}\right] = 0\,,
\ee
derived by imposing the condition that ${\cal D} = {\cal V} = {\cal A} = 0$.
Given our focus on collapse solutions, we plot the negative roots of 
Eq. (57) in Figure 14 as a function of $a$ for values of $C_1
= 1.2, 1.5, 1.8, 2.1$ and $2.4$.  

Solutions that pass through the critical curve can be
obtained by numerically integrating inward and outward
from both sides, using the analytical solutions
obtained via the Taylor series expansion presented in \S6.1 (where
$v_c = k x_c$).  We note that the values of $v_1$ and $\alpha_1$ 
obtained by solving Eqs. (43) - (55) are now independent of $x_c$ -- 
a consequence of the ODE's 
which describe the gas dynamics (Eqs. 17) being invariant
to the scaling transformation $x \rightarrow \eta x$,
$v\rightarrow \eta v$, $\alpha\rightarrow\alpha$, 
$p\rightarrow \eta^2 p$, and $m\rightarrow \eta^2 m$
for the case $\gamma = 1$.
We therefore set $x_c = 1$ in our analysis
without loss of generality.  

Through numerical exploration, we find a class of solutions
that have the limiting form $\alpha \propto x^{2/a}$ and
$v\propto x^{(1+a)/a}$ in the limit $x\to \infty$, and
$\alpha\propto x^{-2/C_1}$ and $v\propto x^{(2-C_1)/C_1}$
in the limit $x\to 0$, where $1 < C_1 < 2$.
To illustrate this type of collapse flow, we plot solutions for the 
case $a = -1.05 $ and $C_1 = 1.5$ in Figure 15. 
The upper panel presents the reduced density profile, and the
lower panel presents the reduced velocity profile. The dotted curves
in these panels represent the ($\alpha_c, x_c$) and ($v_c, x_c$) 
projections of the critical line.  
Solutions for the real density
and real velocity, scaled so that $x = 1$ corresponds to 
$t$ = 1 Myr and $r$ = 1 pc, are presented in Figs. 16 and 17 for times 
0.01, 0.1, 1 and 10 Myrs.   Clearly, this case represents a class of collapse solutions
characterized by a nonzero initial inward velocity
that slows with time at a fixed radius after the gas becomes isothermal.\footnote{
This class of solutions is analogous to the Type II solutions presented in Lou \& Cao (2007)
for a spherical geometry with $\gamma = 4/3$}.  Indeed, the gas at large radii/early times
($x \to \infty$) is moving inward with nearly constant velocity, though the magnitude of the velocity
differs from one part of the flow to another (i.e., each "parcel" of gas
moves with a nearly constant velocity, such that $du/dt \approx 0$, but the velocity field 
for the gas depends on $r$ and $t$).   The density increases 
as the gas falls inward ($x \to 0$), leading to a rise in pressure that
eventually becomes effective at
slowing the inward collapse. The break in density at $x\sim 1$ in this case
results from the slowing down of the infalling gas, rather than from
an inward-out collapse associated with typical self-similar collapse solutions.
This class of solutions could, in part, idealize the dynamics of a cylindrical
structure collapsing after the loss of magnetic support via ambipolar 
diffusion, where it becomes isothermal as turbulence dissipates during the collapse.  

\section{CONCLUSION}  
The main goal of our work was to obtain 
self-similar solutions which describe the collapse
of an initially stationary, cylindrically symmetric gas that is overdense
from hydrostatic equilibrium by a factor $\Lambda$.  Hydrostatic equilibrium 
profiles are easily found for an assumed static equation of state
$p = \alpha^\Gamma$, as specified through the choice of the index $\Gamma$.
Unlike previous works, we allow the equation of state to evolve during
the collapse, under the condition that entropy is conserved along a streamline.
Doing so allows the equation of state to change from its initial form (as
defined by $\Gamma$), to a different polytropic form as defined by the dynamic
index $\gamma$.  Physical solutions describing this type of collapse
require $0 < \Gamma < 1$ and $\gamma\ne 1$.  
We present solutions for which the system 
equations do not become singular, as well as solutions
which pass smoothly through the singular surface.

Our solutions
clearly exhibit the broken power law profiles 
typical of self-similar collapse flows.  We find that the spectral
indices which characterize the reduced density solutions are not sensitive
to the choice of $\gamma$, and hence have similar forms when
the similarity variable $x = At^a r << 1$.
In contrast, these solutions depend sensitively on the value of $\Gamma$, both
in terms of shape when $x >> 1$ and in overall normalization. 
Similar behavior is also clearly observed for the reduced velocity solutions.  
The insensitivity of the collapse dynamics to 
the value of $\gamma$ results from 
the inside-out nature of the collapse.  
Specifically, the break at $x \approx 1$ exhibited in the density
and velocity profiles (as shown in Figs. 4 -- 7) occurs as a result
of the gas being in some part supported by 
pressure ($x >> 1$) evolving to a state approaching free-fall ($x << 1$)
as a result of the loss of that pressure support.  
This break-point also denotes a  
transition from an initial (static) equation of state
($p = \alpha^\Gamma$) to a dynamic equation of state
($p \propto \alpha^\gamma$), with the 
transition occurring within a narrow region around $x \approx 1$. 
In terms of real variables, this transition occurs at a boundary 
which moves outward through the gas as governed by the static
equation of state.  As such, the gas dynamics depend sensitively on 
the value of $\Gamma$
(which, of course, also governs the density profile $\rho(r)$ of the
initial state).
In contrast, the gas pressure within this boundary, as
described by the dynamic equation of state (and in turn $\gamma$), has little effect
on the overall collapse dynamics.

Although self-similar collapse flows  
have no characteristic mass scale, the
collapse flow feeds material onto the central star/disk system at a
well-defined mass infall rate $\dot M$. Since the infalling
material always approaches free-fall conditions on the inside (in the
limit $r \to 0$), the reduced mass determines the size of the
infall rate, and is therefore an important parameter in 
the collapse problem. Our analysis shows that, as expected,
larger infall rates
occur for initial states that are more overdense.
In addition, stiffer static equations of state result in larger mass
accretion rates.
In contrast, the dynamic equation of state (as defined
by $\gamma$) has little effect on the value of $m_0$.

Our results indicate that collapse from an overdense state initially at rest cannot
occur if $\gamma = 1$, regardless of the form for the initial equation
of state.  This result is somewhat surprising
given the apparent insensitivity to $\gamma$ exhibited
by all other solutions we obtain.  However, we do find a different class of collapse 
solutions which passes smoothly
through the singular surface when $\gamma = 1$.  These solutions are
analogous to the Type II solutions presented in Lou \& Cao (2007)
for a spherical geometry with $\gamma = 4/3$.

\bigskip 
{} 
\bigskip 
%\newpage 
\centerline{\bf Acknowledgments} 

We thank the anonymous referee for useful comments.  We also thank Fred Adams for many useful discussions. 
BB was supported by the Greaves Fund at Northern Kentucky 
University. MF was supported by the
Hauck Foundation at Xavier University.

%\bigskip 
\newpage 
\centerline{\bf REFERENCES} 
\medskip 

\par\pp
Adams, F. C. 1990, ApJ, 363, 578

\par\pp
Adams, F. C., Lada, C. J., \& Shu, F. H. 1987, ApJ, 321, 788 

\par\pp
Allen, A., Li, Z.-Yi., \& Shu, F. H. 2003, ApJ, 599, 363

\par\pp
Allen, A., Shu, F. H., \& Li, Z.-Yi. 2003, ApJ, 599, 351 

\par\pp
Cassen, P., \& Moosman, A. 1981, Icarus, 48, 353

\par\pp
Curry, C., \& McKee, C. F. 2000, ApJ, 528, 734 

\par\pp
Fatuzzo, M., Adams, F. C., \& Myers, P. C. 2004, ApJ, 615, 813

\par\pp
Foster, P. N., \& Chevalier, R. A. 1993, ApJ, 416, 303 

\par\pp
Galli, D., \& Shu, F. H. 1993a, ApJ, 417, 220

\par\pp
Galli, D., \& Shu, F. H. 1993b, ApJ, 417, 243

\par\pp
Hennebelle, P. 2003, A\&A, 397, 381

\par\pp
Harjunpaa, P., Kaas, A. A., Carlqvist, P., \& Gahm, G. F. 1999, A\&A, 349, 912

\par\pp
Houlahan, P., \& Scalo, J. M. 1992, ApJ, 393, 172

\par\pp
Hunter, C. 1977, ApJ, 218, 834 

\par\pp
Inustsuka, S., \& Miyama, S. M. 1992, ApJ, 388, 392

\par\pp
Jappsen, A. K., Klessen, R. S., Larson, R. B., Li, Y., \& MacLow, M. M. 2005, 
A\&A, 435, 611

\par\pp
Jijina, J., Myers, P. C., \& Adams, F. C. 1999, ApJ Suppl., 125, 161  

\par\pp
Kawachi, T., \& Hanawa, T. 1998, PASJ, 50, 577

\par\pp
Lada, C. J., Muench, A. A., Rathborne, J., Alves, J. F.,
\& Lombardi, M. 2008, ApJ, 672, 410

\par\pp
Larson, R. B. 1969a, MNRAS, 145, 271

\par\pp
Larson, R. B. 1969b, MNRAS, 145, 297

\par\pp
Larson, R. B. 1981, MNRAS, 194, 809

\par\pp
Li, Z.-Y., \& Shu, F. H. 1996, ApJ, 472, 211L 

\par\pp
Li, Z.-Y., \& Shu, F. H. 1997, ApJ, 475, 237 

\par\pp
Lou, Y.-Q., \& Cao, Y. 2008, MNRAS, 384, 611
 
\par\pp
Penston, M. V. 1969a, MNRAS, 144, 425 

\par\pp
Penston, M. V. 1969b, MNRAS, 145, 457 

\par\pp
Shadmehri, M. 2005, MNRAS, 356, 1429
 
\par\pp
Shu, F. H. 1977, ApJ, 214, 488 

\par\pp
Shu, F. H., Adams, F. C., \& Lizano, S. 1987, A R A \& A, 
25, 23

\par\pp
Stahler, S. W., Shu, F. H., \& Taam R. E. 1980, ApJ, 241, 637  

\par\pp
Terebey, S., Shu, F. H., \& Cassen, P. 1984, ApJ, 286, 529

\par\pp
Tilley, D. A., \& Pudrit, R. E. 2003, ApJ, 593, 426

\par\pp
Ulrich, R. K. 1976, ApJ, 210, 377

\par\pp
Whitworth, A., \& Summers, D. 1985, MNRAS, 214, 1

\newpage 
\begin{figure}
\figurenum{1}
{\centerline{\epsscale{0.90} \plotone{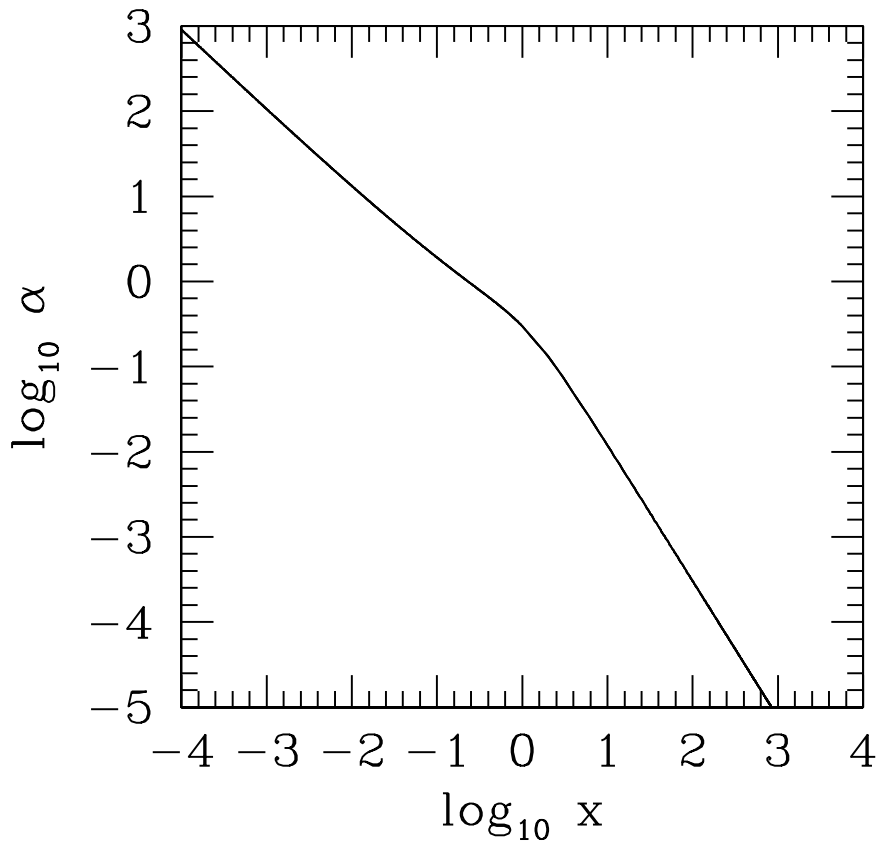} }}
\figcaption{Self-similar solutions for the reduced density 
describing the collapse of a gas defined by 
the parameters $a = -1.25$ and $\gamma = 0.25$.
All solutions are integrated inward from an initial 
set of values $x_i = 5 \times 10^3$ and $\alpha(x_i) = 5.78
\times 10^{-7}$, and the five values of
$v_i$ as discussed in the text.  
All five solutions are within $0.1$ percent of each other,
illustrating that the self-similar density profiles are
not sensitive to the initial values of the reduced velocity. 
}  
\end{figure}

\newpage 
\begin{figure}
\figurenum{2}
{\centerline{\epsscale{0.90} \plotone{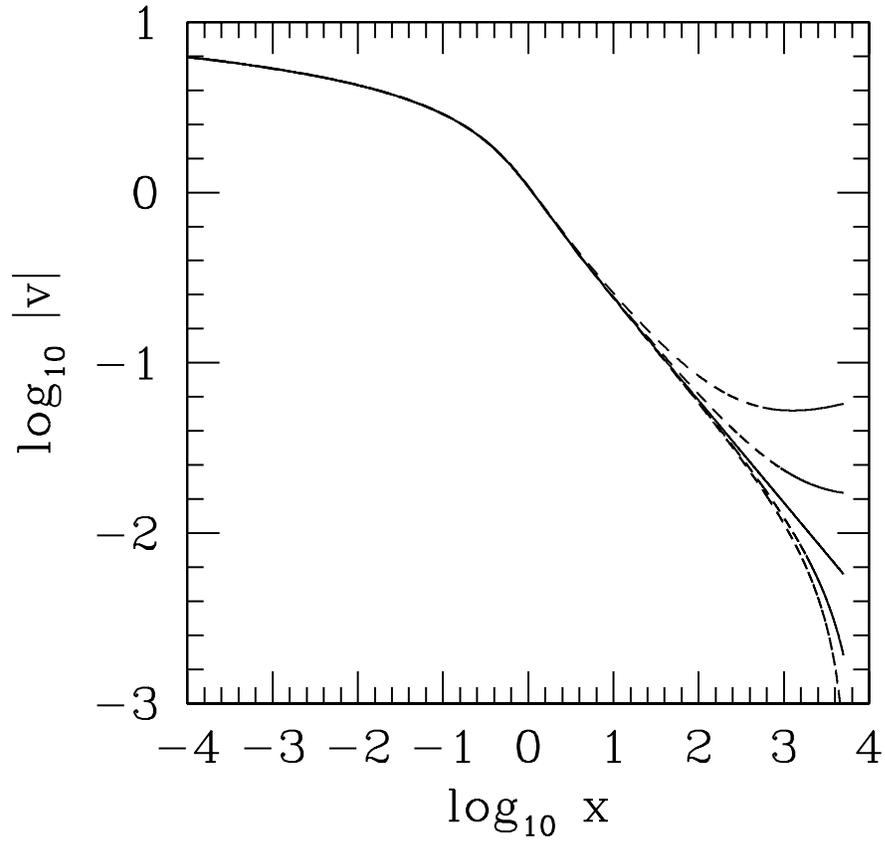} }}
\figcaption{Self-similar solutions for the reduced velocity 
describing the collapse of a gas defined by 
the parameters $a = -1.25$ and $\gamma = 0.25$.
All solutions are integrated inward from an initial 
set of values $x_i = 5 \times 10^3$ and $\alpha(x_i) = 5.78
\times 10^{-7}$, and the five values of
$v_i$ as discussed in the text.  
}  
\end{figure}

\newpage 
\begin{figure}
\figurenum{3}
{\centerline{\epsscale{0.90} \plotone{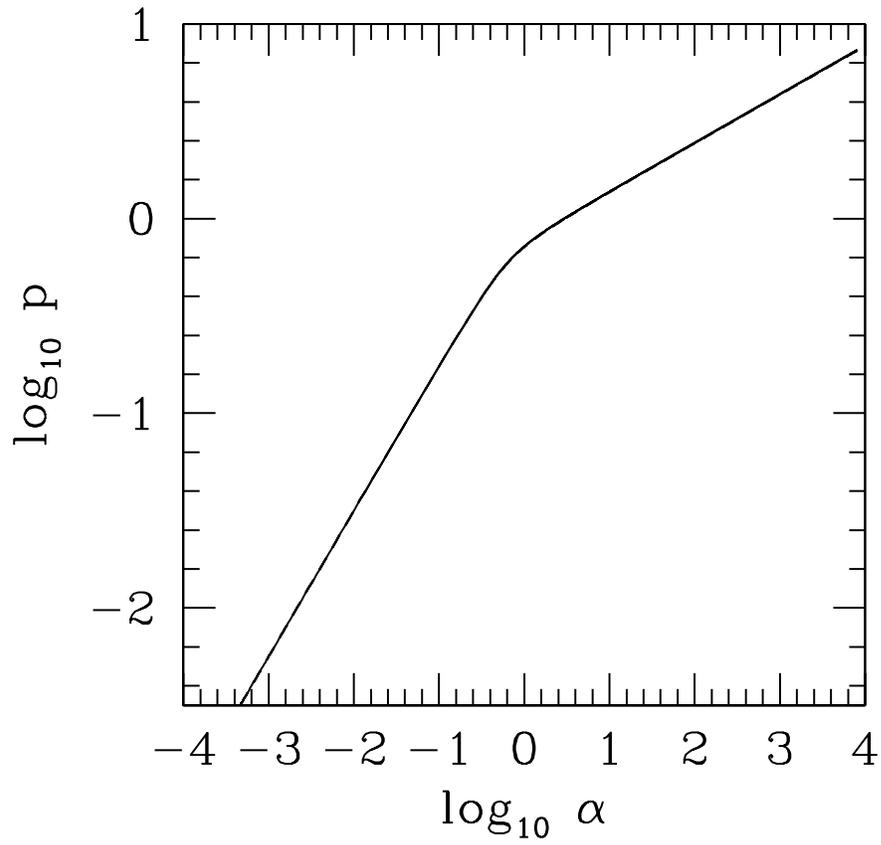} }}
\figcaption{Log $p$ - log $\alpha$ profiles for
the solutions presented in Figs. (1) and (2), 
illustrating the transition during the collapse 
from a static equation 
of state ($p = \alpha^{a+2} = \alpha^{0.75}$) as shown in the
lower left corner to a dynamic
equation of state ($ p \propto\alpha^\gamma \propto\alpha^{0.25}$) 
as shown in the upper right corner.
}  
\end{figure}

\begin{figure}
\figurenum{4}
{\centerline{\epsscale{0.90} \plotone{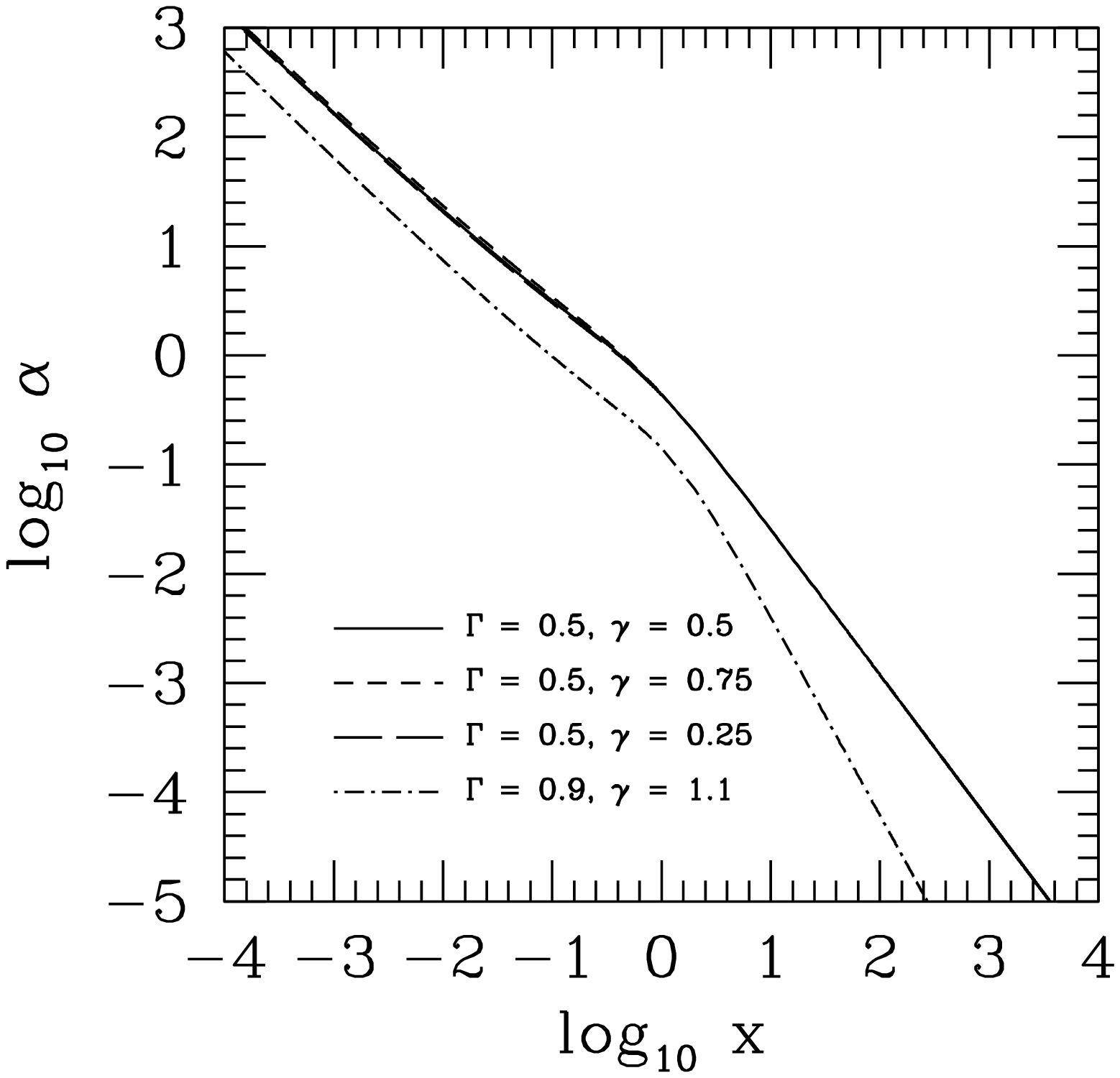} }}
\figcaption{Self-similar solutions for the reduced density 
describing the collapse of an initially static
gas overdense from its hydrostatic
equilibrium state by a factor of $\Lambda = 1.5$.
The four profiles correspond to the four different
choices of static and dynamics equations of state,
as defined by the specified values of $\gamma$
and $\Gamma$.
}  
\end{figure}
\newpage 
\begin{figure}
\figurenum{5}
{\centerline{\epsscale{0.90} \plotone{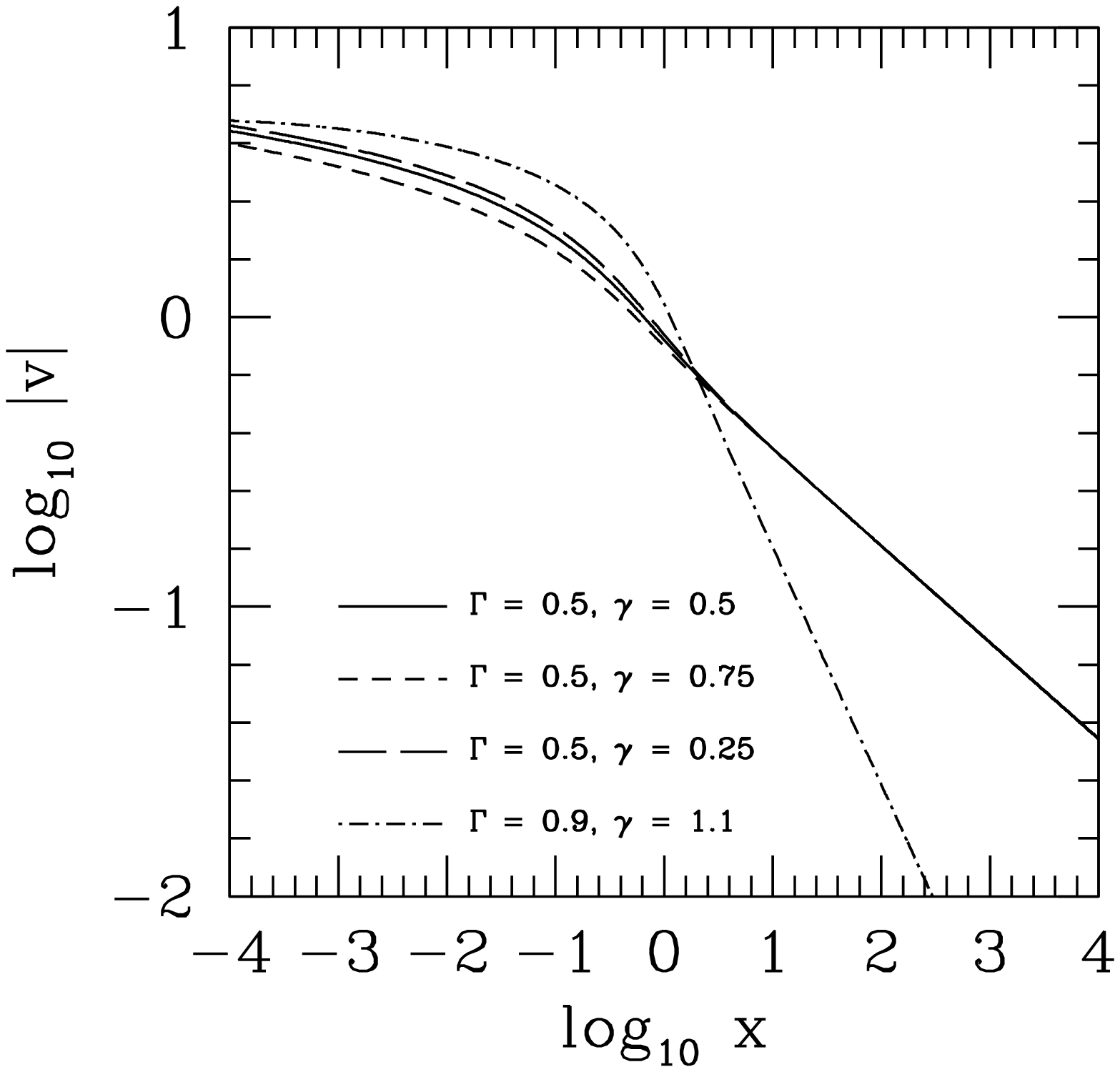} }}
\figcaption{Self-similar solutions for the reduced velocity
describing the collapse of an initially static
gas overdense from its hydrostatic
equilibrium state by a factor of $\Lambda = 1.5$.
The four profiles correspond to the four different
choices of static and dynamics equations of state,
as defined by the specified values of $\gamma$
and $\Gamma$.
}  
\end{figure}

\newpage 
\begin{figure}
\figurenum{6}
{\centerline{\epsscale{0.90} \plotone{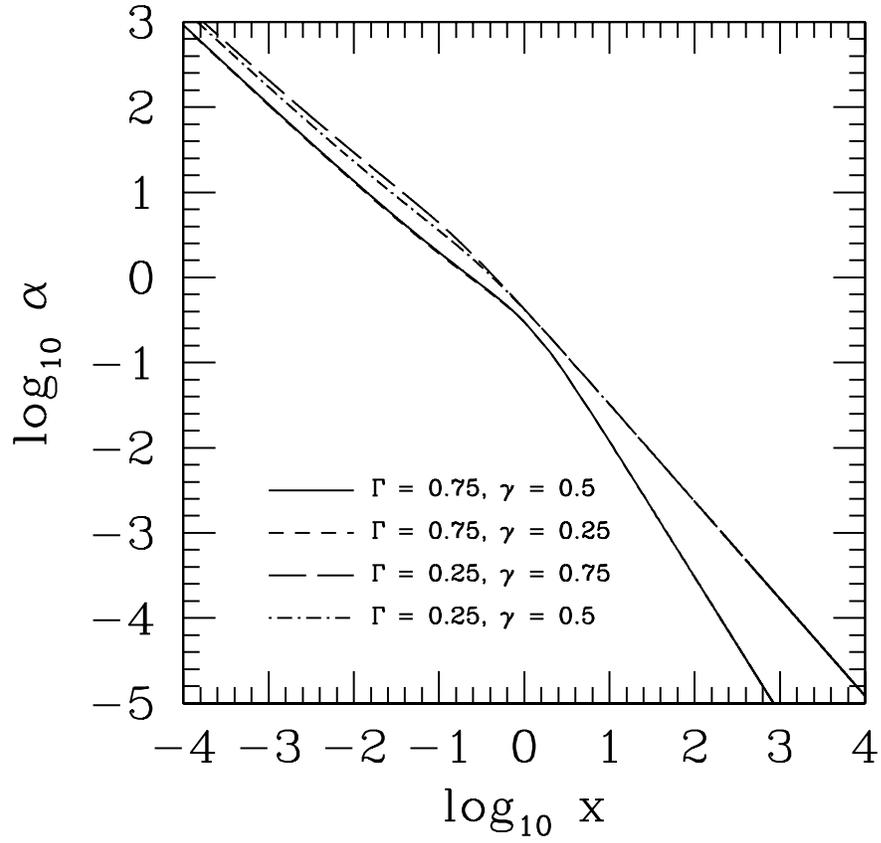} }}
\figcaption{Same as Figure 4, but for a different set
of values for $\gamma$ and $\Gamma$. We note that the
solid curve and short-dashed curve lie over each other
in this Figure, and are therefore hard to distinguish.
}  
\end{figure}

\newpage 
\begin{figure}
\figurenum{7}
{\centerline{\epsscale{0.90} \plotone{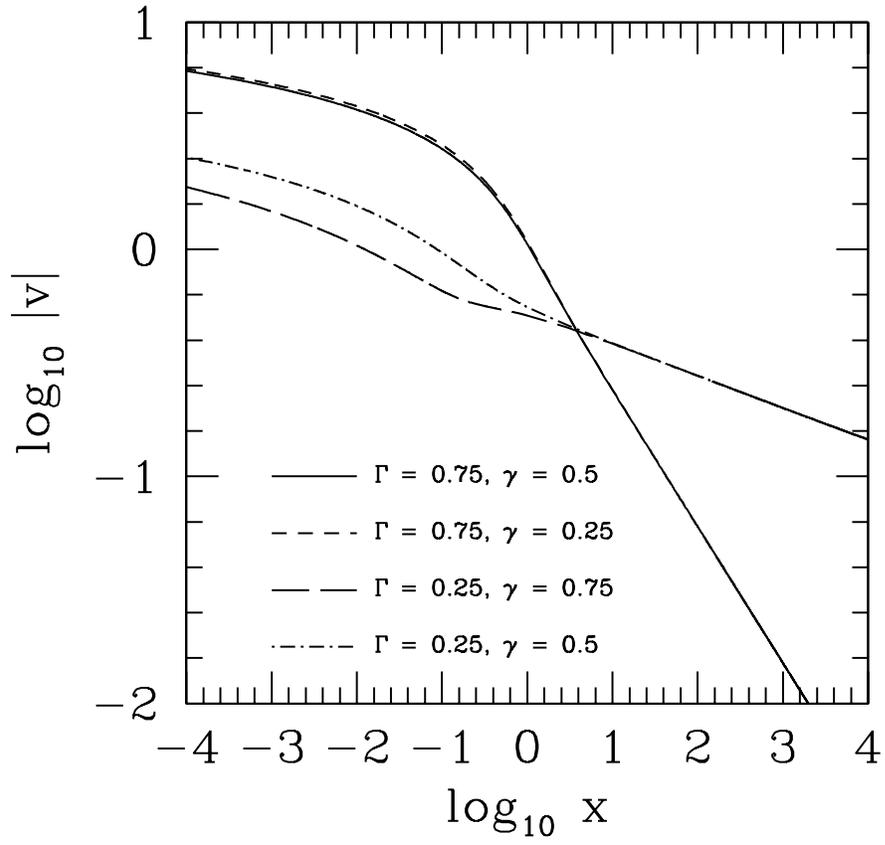} }}
\figcaption{Same as Figure 5, but for a different set
of values for $\gamma$ and $\Gamma$.  
}  
\end{figure}

\newpage 
\begin{figure}
\figurenum{8}
{\centerline{\epsscale{0.90} \plotone{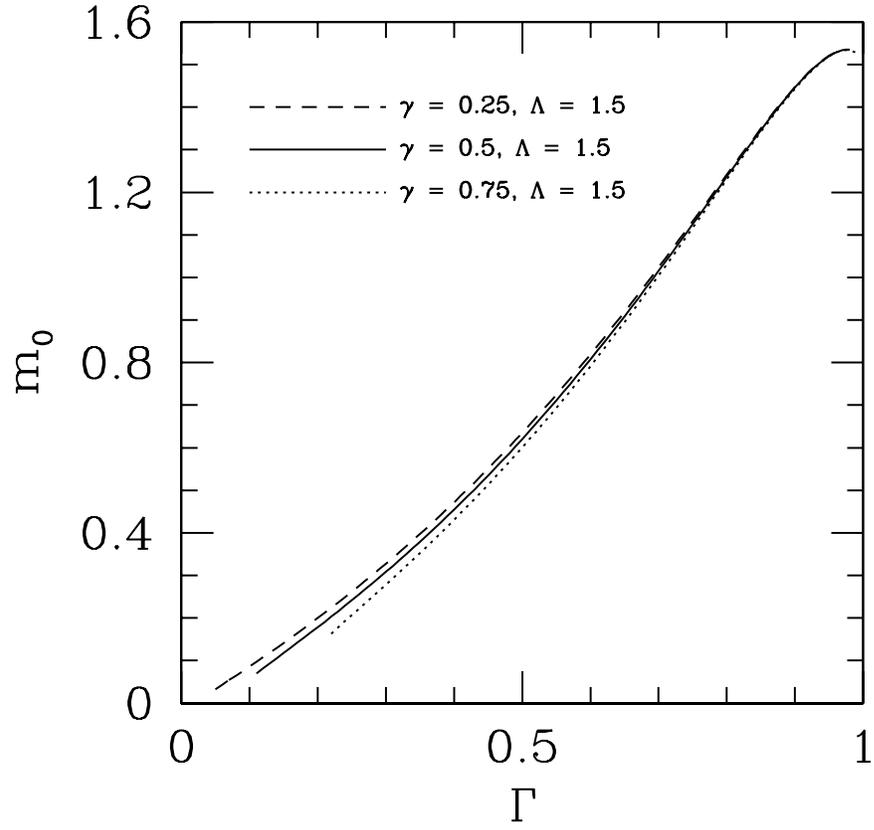} }}
\figcaption{The value of the reduced mass at the origin
$m_0$ as a function of the static equation of state
index $\Gamma$ for $\Lambda = 1.5$ and three different values of 
the dynamic equation of state index $\gamma$. 
}  
\end{figure}

\newpage 
\begin{figure}
\figurenum{9}
{\centerline{\epsscale{0.90} \plotone{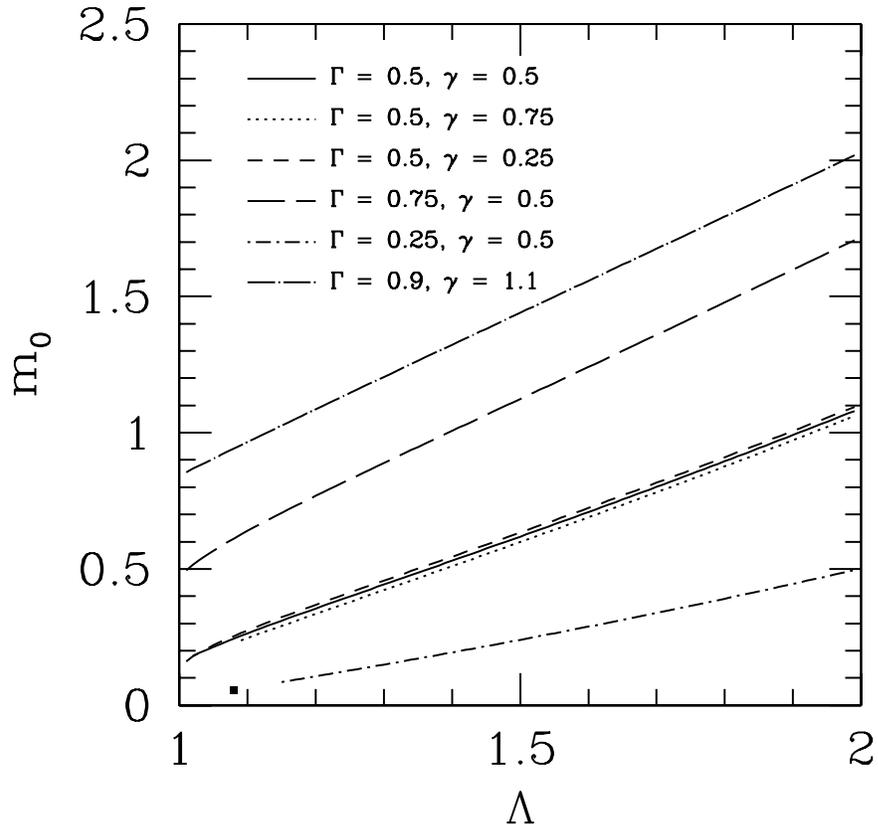} }}
\figcaption{The value of the reduced mass at the origin
$m_0$ as a function of the overdensity 
parameter $\Lambda = 1.5$ for the specified values of
the equation and state indices $\Gamma$ and $\gamma$.
The solid square data point corresponds to the results
obtained for a solution which passes smoothly through the
singular surface, as discussed in \S5.1.
}  
\end{figure}

\newpage 
\begin{figure}
\figurenum{10}
{\centerline{\epsscale{0.90} \plotone{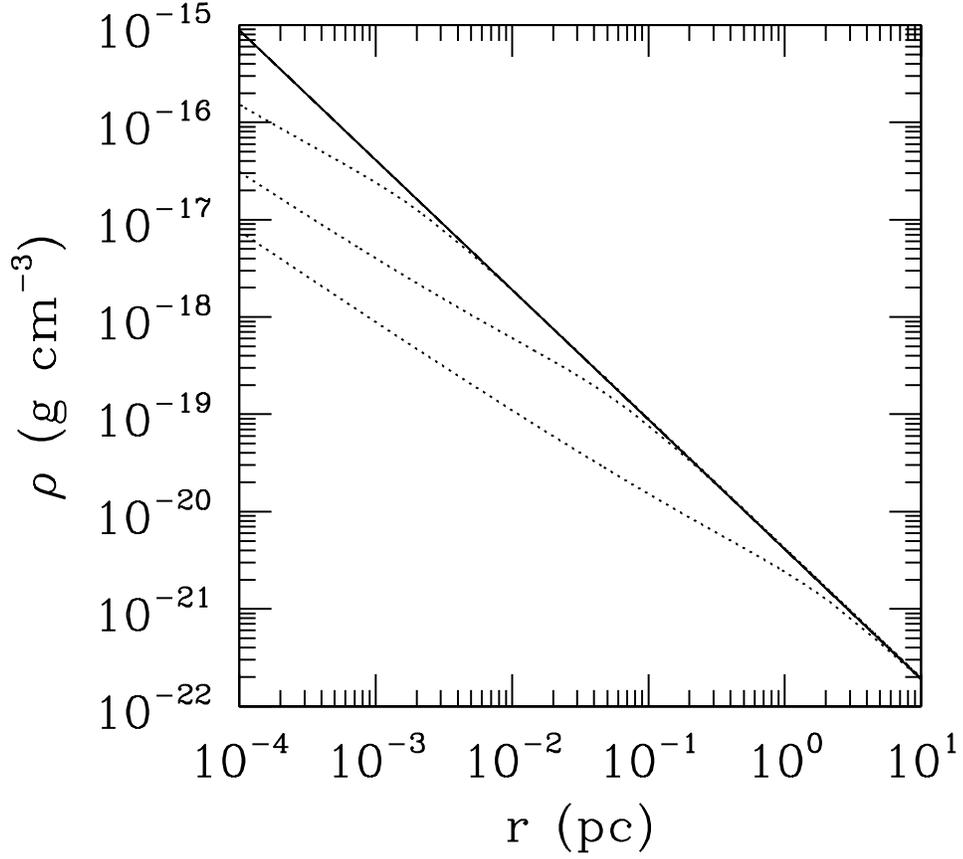} }}
\figcaption{Density profiles illustrating the inside-out collapse
of a cylindrical cloud initially at rest and overdense by a
factor $\Lambda  = 1.5$, and for which $\Gamma = \gamma = 0.5$. 
The solid curve represents the initial ($t = 0$) density profile
$\rho = \Lambda \rho_E$, 
and the dotted curves show the profiles (from top to 
bottom) at times $t = 10^4$, $t = 10^5$ and 
$t = 10^6$ years.  
}  
\end{figure}

\newpage 
\begin{figure}
\figurenum{11}
{\centerline{\epsscale{0.90} \plotone{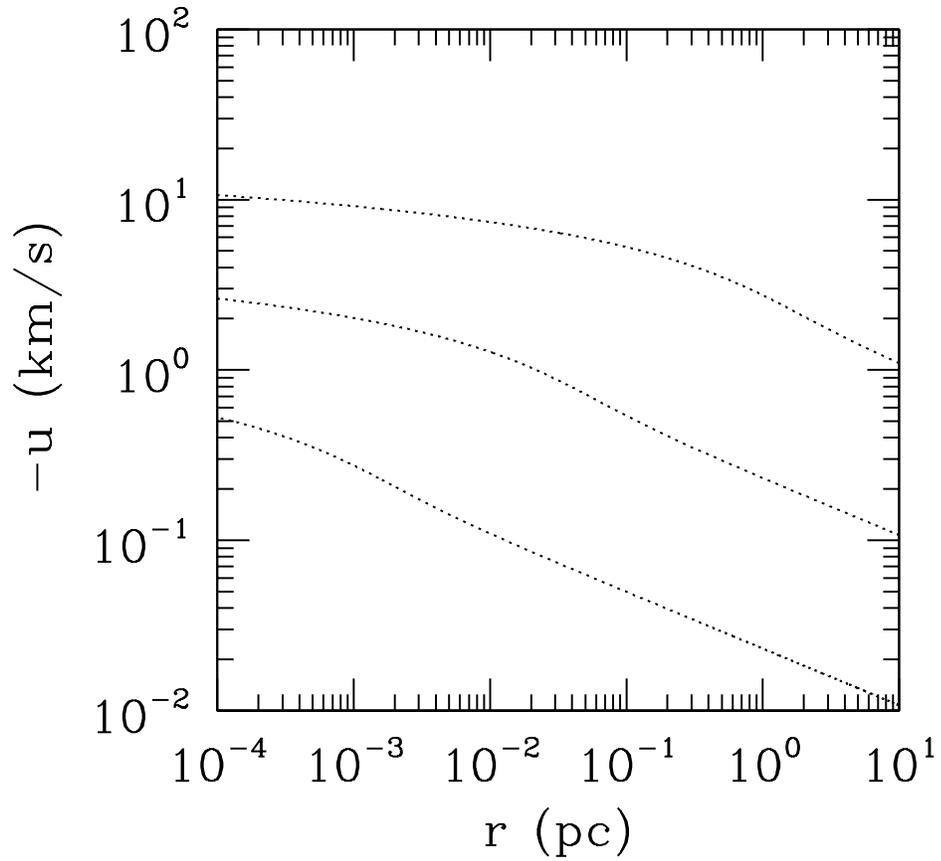} }}
\figcaption{Velocity profiles illustrating the inside-out collapse
of a cylindrical cloud initially at rest and overdense by a
factor $\Lambda  = 1.5$, and for which $\Gamma = \gamma = 0.5$. 
The dotted curves show the profiles (from bottom to 
top) at times $t = 10^4$, $t = 10^5$ and 
$t = 10^6$ years.  
}  
\end{figure}

\newpage 
\begin{figure}
\figurenum{12}
{\centerline{\epsscale{0.90} \plotone{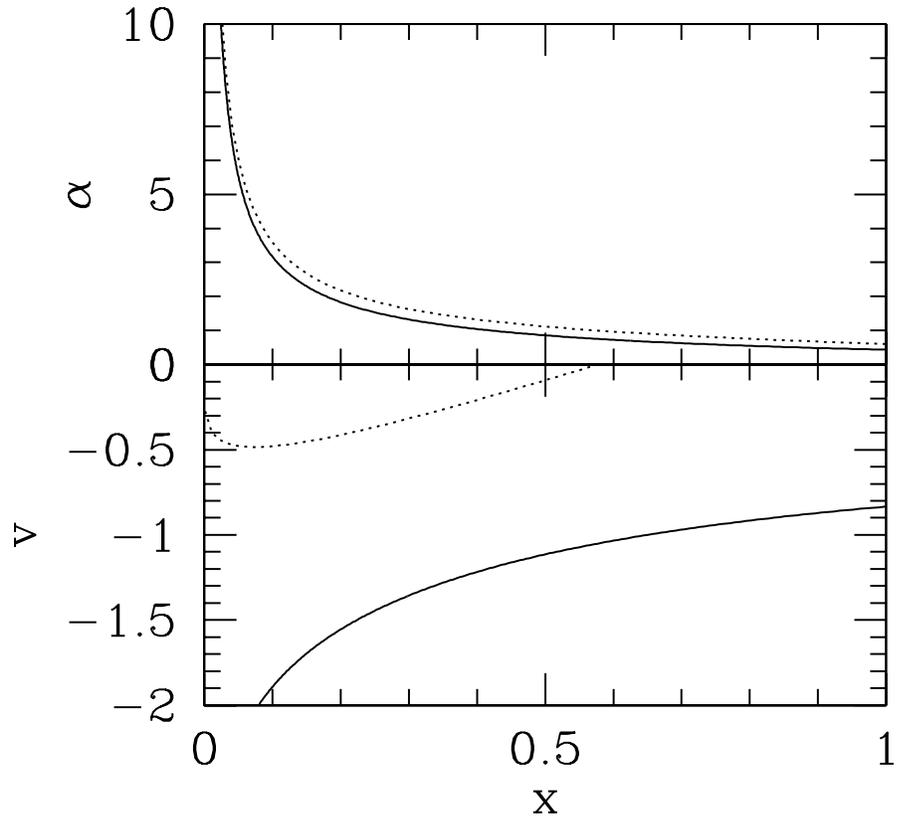} }}
\figcaption{The dotted curves denote the density (upper panel) 
and velocity (lower panel)
projections of the critical curve for the parameters  
$\Gamma = \gamma = 0.5$ and $\Lambda = 1.5$. The solid curve in each panel
denotes the corresponding self-similar solution, also 
shown in Figs. 4 and 5.  This solution does not
pass through the critical curve.
}  
\end{figure}

\newpage 
\begin{figure}
\figurenum{13}
{\centerline{\epsscale{0.90} \plotone{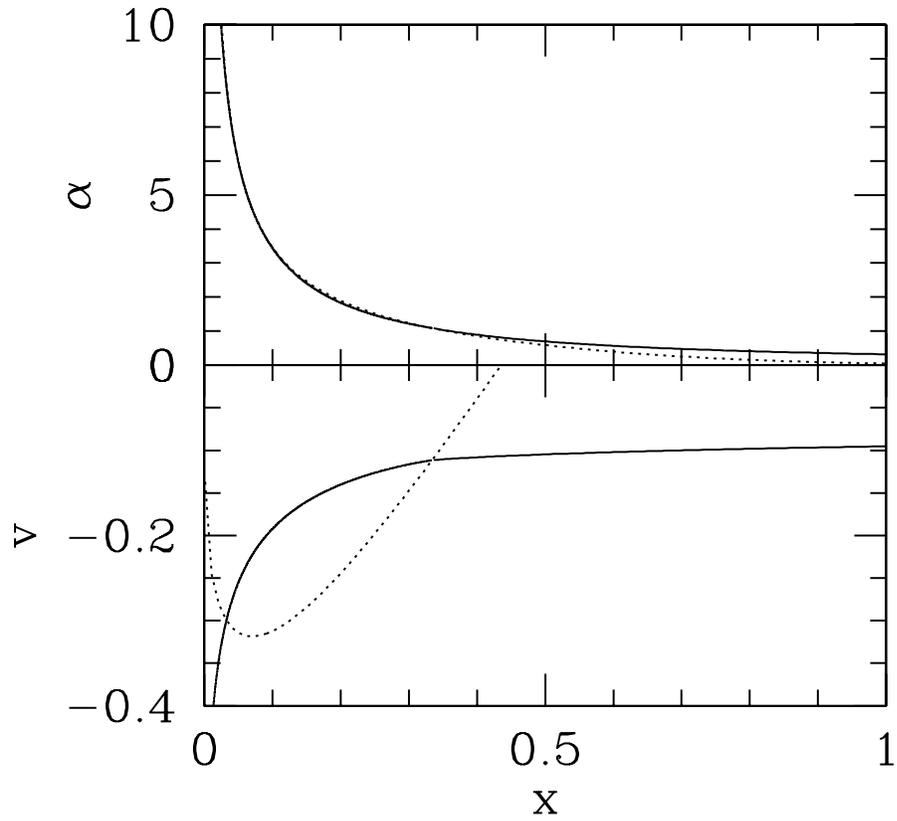} }}
\figcaption{The dotted curves denote the density (upper panel) 
and velocity (lower panel)
projections of the critical curve for the parameters  
$\Gamma = 0.25$, $\gamma = 0.5$, and $\Lambda = 1.08$. The solid curve in each panel
denotes the corresponding self-similar solution, which
passes through the critical curve at
$x_c = 0.336$ and at $x_c = 0.0322$.
}  
\end{figure}

\newpage 
\begin{figure}
\figurenum{14}
{\centerline{\epsscale{0.90} \plotone{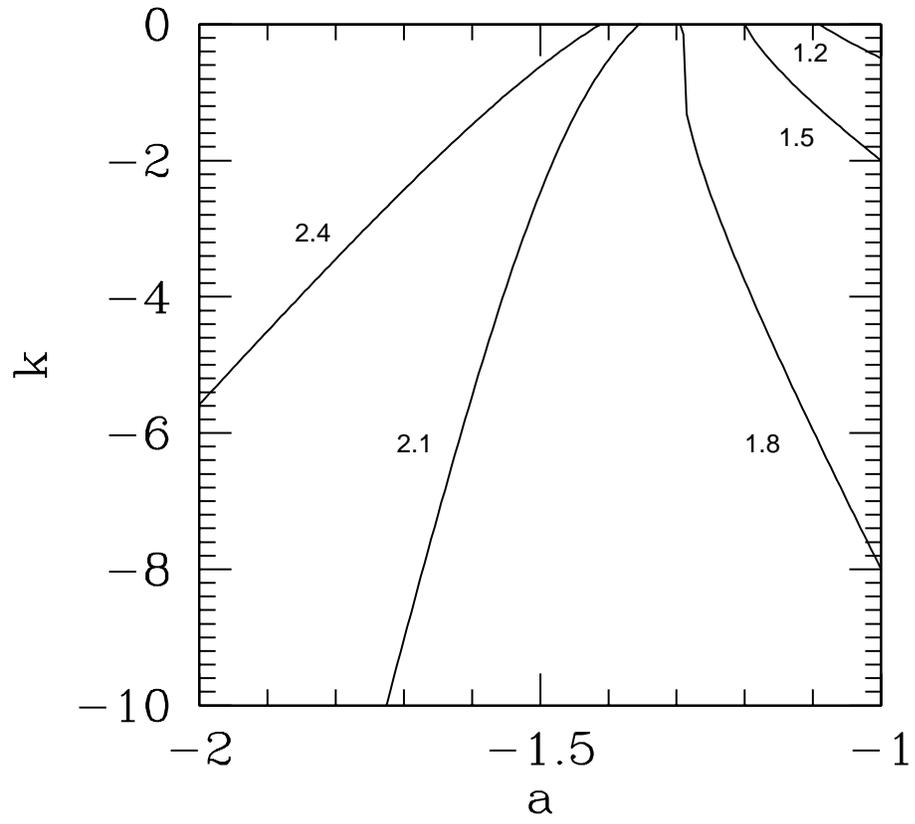} }}
\figcaption{The negative (real) roots of
Eq. xxxxxx as a
function of $a$ for values of 
$C_1 = 1.2, 1.5, 1.8, 2.1$ and $2.4$, as denoted
in the figure. 
}  
\end{figure}

\newpage 
\begin{figure}
\figurenum{15}
{\centerline{\epsscale{0.90} \plotone{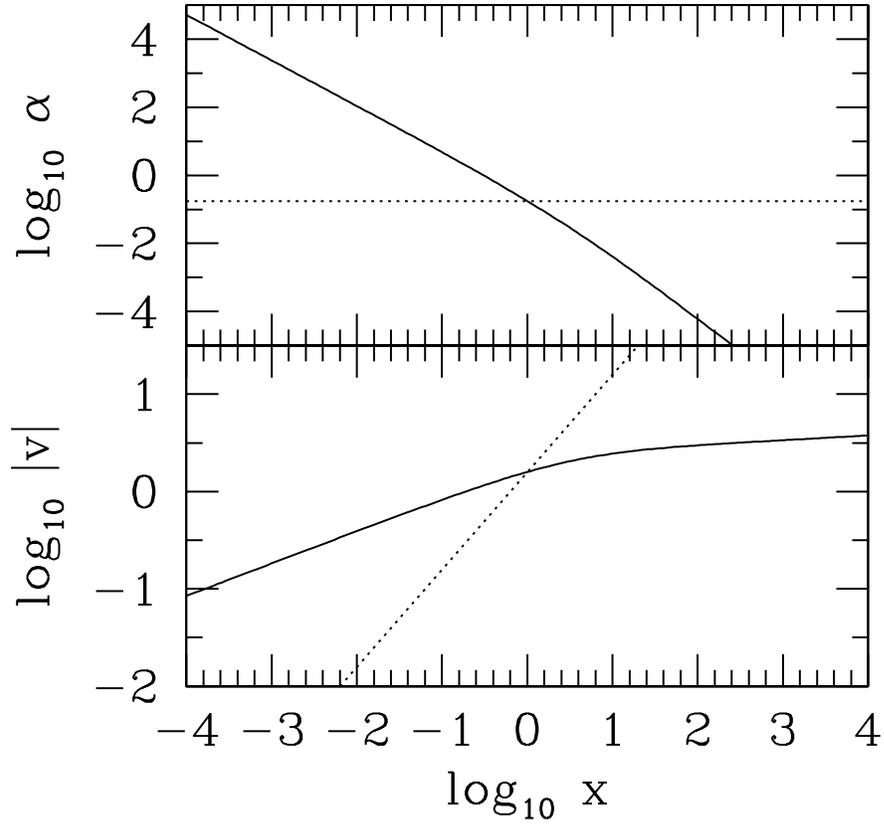} }}
\figcaption{The dotted curves denote the density (upper panel) 
and velocity (lower panel)
projections of the critical line for the parameters  
$a = -1.05$, $C_1 = 1.5$. The solid curve in each panel
denotes the corresponding self-similar solution which passes through
the critical line at the point $x_c = 1$. 
}  
\end{figure}

\newpage 
\begin{figure}
\figurenum{16}
{\centerline{\epsscale{0.90} \plotone{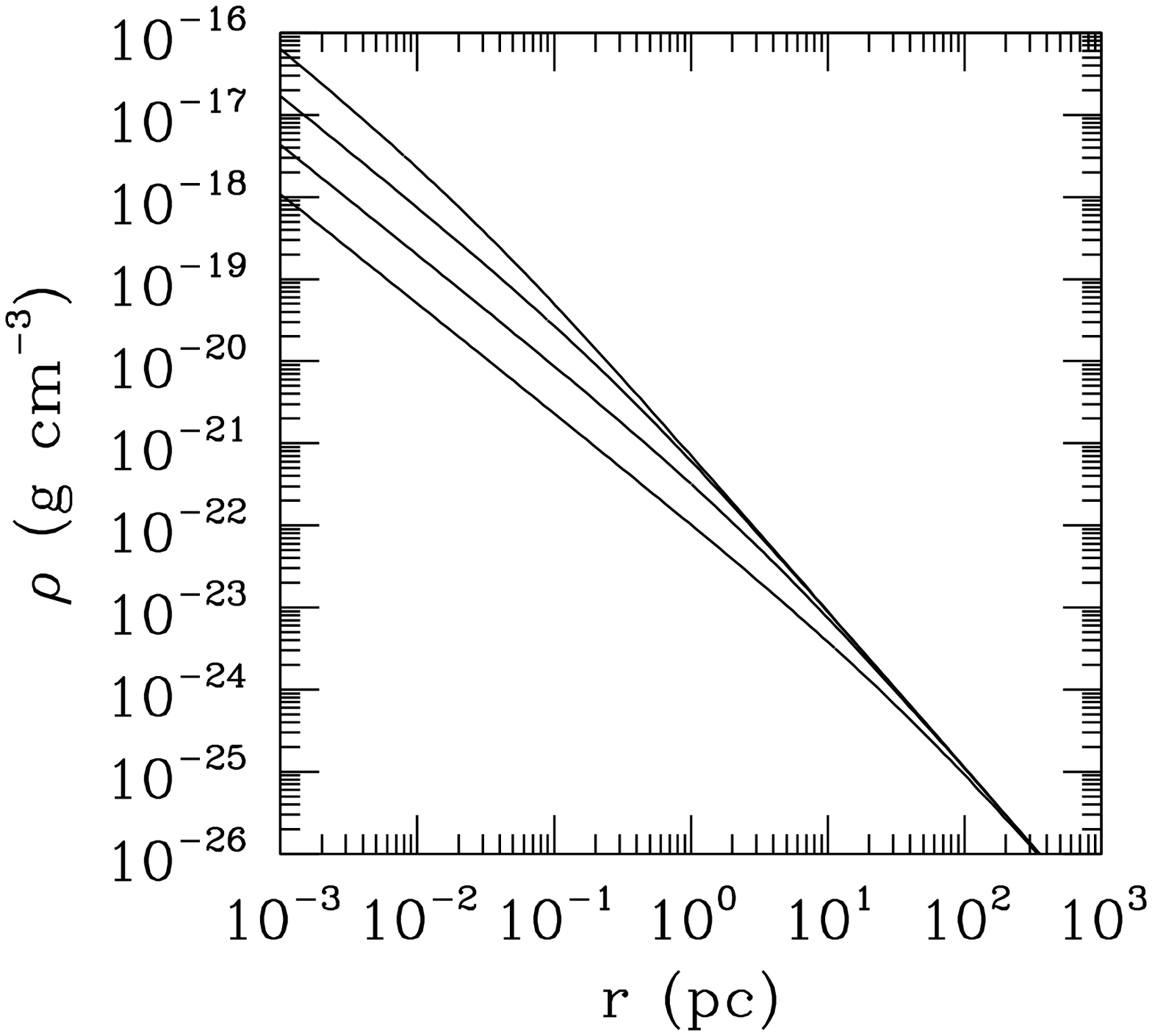} }}
\figcaption{Density profiles associated with the self-similar solutions
presented in Fig. 15, scaled by setting $x_c = 1$ when
$r = 1$ pc and $t = 1$ Myr.  The solid curves show the profiles (from top to 
bottom) at times  $t = 0.01$, $0.1$, $1$ and $10$ Myrs.
}  
\end{figure}

\newpage 
\begin{figure}
\figurenum{17}
{\centerline{\epsscale{0.90} \plotone{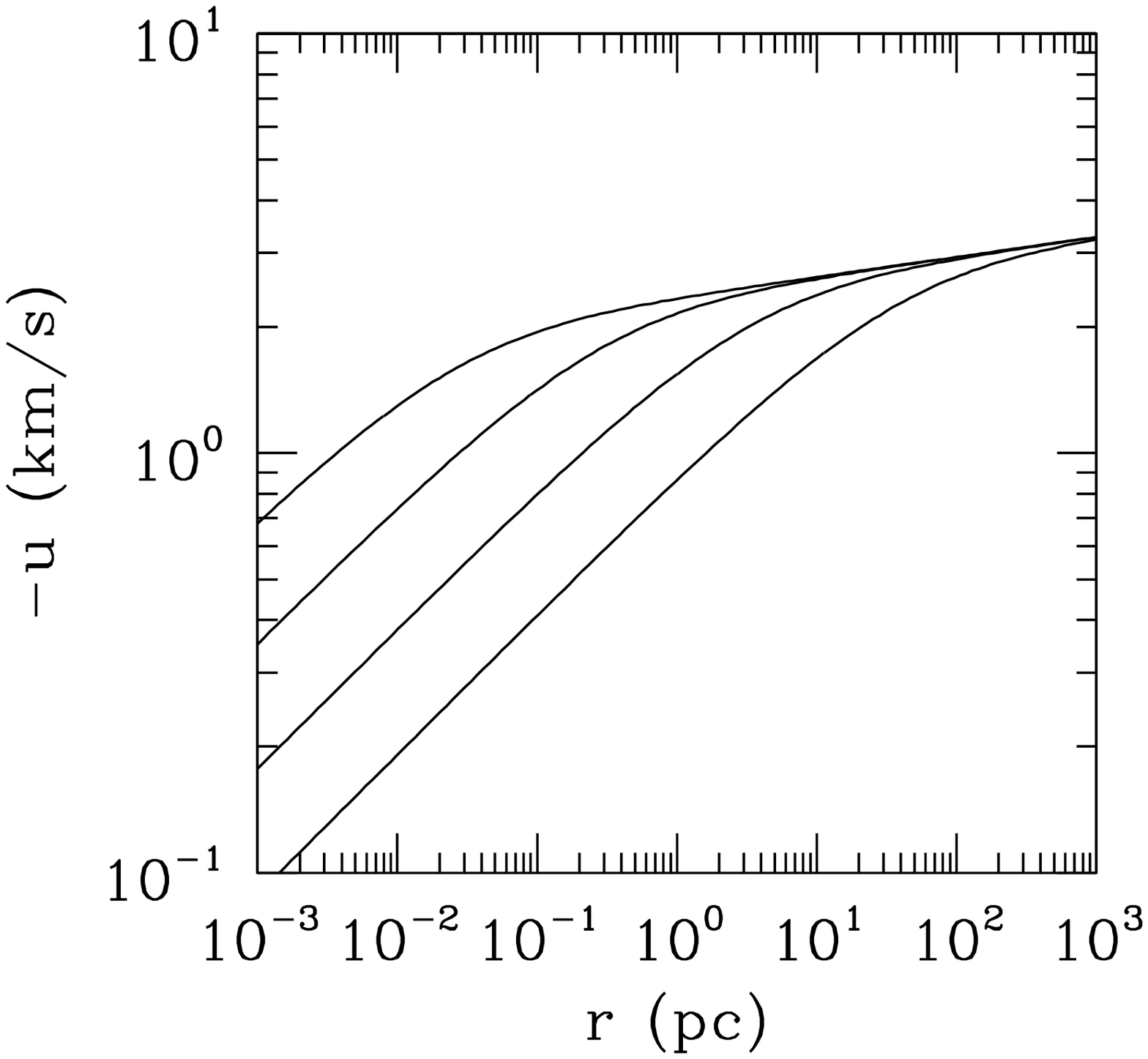} }}
\figcaption{Velocity profiles associated with the self-similar solutions
presented in Fig. 15, scaled by setting $x_c = 1$ when
$r = 1$ pc and $t = 1$ Myr.  The solid curves show the profiles (from top to 
bottom) at times $t = 0.01$, $0.1$, $1$ and $10$ Myrs.
}  
\end{figure}

\end{document}